\input harvmac.tex

\def\exp{{\rm exp}}

\def\sin{{\rm sin}}
\def\cos{{\rm cos}}

\def\frac#1#2{{#1\over#2}}

\input epsf.tex

\Title{\vbox{\baselineskip12pt\hbox{CLNS 96/1400}
                              \hbox{RIMS-1063}
                              \hbox{hep-th/9602074}}}
{\vbox{\centerline{
Multi-point Local Height Probabilities}
\vskip6pt\centerline{ in the Integrable RSOS Model}}}

\centerline{Sergei Lukyanov$^{1,2,3}
$\footnote{$^*$}
{e-mail address: sergei@hepth.cornell.edu}
\ and Yaroslav Pugai$^{1,3}$
\footnote{$^{**}$}
{e-mail address: slava@itp.ac.ru}}
\bigskip
\centerline{$^1$ RIMS, Kyoto University,
Sakyo-ku, Kyoto 606, Japan}

\centerline{$^2$\ Cornell University, Ithaca, NY 14853-5001, USA}

\centerline{and}

\centerline{$^3$\ L.D. Landau Institute for Theoretical
Physics}
\centerline{Chernogolovka, 142432, Russia}

\bigskip\bigskip\bigskip
\centerline{\bf{Abstract}}
\bigskip
By using the bosonization technique,
we derive  an
integral representation for multi-point
Local Hight Probabilities
for the  Andrews-Baxter-Forrester model
in the regime III.
We argue  that
the dynamical symmetry of the model  is
provided by the deformed Virasoro algebra.

\Date{February, 96}

\lref\japnews{Jimbo, M. and Miwa, T:
Quantum KZ equation with\  $|q|=1$\ 
and correlation functions of
the XXZ model in the gapless regime.
RIMS preprint RIMS-1058 (1996)
(hep-th/9601135)}

\lref\CFT{Itzykson, E., Saleur, H. and Zuber, J.B. (eds.): Conformal
Invariance and Applications to
Statistical Mechanics, World Scientific (1988)}

\lref\cristal{Domb, C. and Lebowitz, J.L. (eds):
Phase transitions and Critical Phenomena,
vol. {\bf 12}, Academic Press, New York  (1988) }

\lref\singordon{Lukyanov, S.: Free Field Representation for Massive
Integrable Models,
Commun. Math. Phys. {\bf 167}, 1, 183-226 (1995)\semi
Lukyanov, S.: Correlators of the Jost
functions  in the Sine-Gordon Model. Phys. Lett.
{\bf B325}, 409-417 (1994)}

\lref\saleur{Saleur, H. and Bauer, M.:
On some relations between local height
probabilities and conformal invariance,
Nucl. Phys. {\bf B320}, 591-624 (1989)}

\lref\note{
Lukyanov, S.:
A note on the deformed Virasoro algebra.
Phys. Lett.
{\bf B367}, 121-125 (1996)}

\lref\las{ Lashkevich, M. and  Pugai, Ya.:
unpublished}

\lref\fr{Frenkel, E. and Reshetikhin, N.:
Quantum affine algebras and deformations of
the Virasoro and W-algebras. Preprint (1995) (q-alg/9505025) }

\lref\Baxeight{Baxter, R.J.: Corner Transfer matrices
of the eight vertex model. Low-temperature
expansions and conjectured properties.
J. Stat. Phys. {\bf  15}, 485-503 (1976)}

\lref\japAn{Date, E., Jimbo, M., Kuniba, A., Miwa, T. 
and  Okado, M.: Exactly 
solvable  SOS models: Local height probabilities and theta functions 
identities. Nucl. Phys.  B{\bf 290}, 231-273, (1987)\semi
Date, E., Jimbo, M., Kuniba, A., Miwa, T. and  Okado, M.: Exactly
solvable  SOS models 2: Proof of the star-triangle
relation and combinatorial identities.
Adv. stud. Pure Math. {\bf 16}, 17-122 (1988)}

\lref\huse{Huse, D.A.:
Exact exponents for infinitely many new multi-critical
points. Phys. Rev. {\bf B30}, 3908-3915 (1984)}

\lref\japtwo{Davies, B., Foda,
O., Jimbo, M., Miwa, T. and
Nakayashiki, A.:
Diagonalization of the XXZ Hamiltonian by
Vertex Operators. Commun. Math. Phys. {\bf 151}, 89-153  (1993)}

\lref\japthree{Jimbo, M., Miki, K., Miwa, T. 
and  Nakayashiki, A.:
Correlation functions of the XXZ model for $\Delta<-1$.
Phys. Lett. {\bf A168}, 256-263 (1992)}

\lref\japising{Foda, O., Jimbo, M., Miwa, T., Miki, K.
and  Nakayashiki, A.:
Vertex operators in solvable lattice models. J. Math. Phys. 
{\bf 35}, 13-46 (1994)}

\lref\cls{Clavelli, L. and  Shapiro, J.A.:
Pomeron Factorization in 
General Dual Models. Nucl. Phys. {\bf B57}, 490-535 (1973)}

\lref\fei{Feigin, B.L. and  Fuchs, D.B.:
Representations
of the Virasoro algebra. In:
Topology, Proceedings, Leningrad 1982.
Faddeev, L.D., Mal'cev, A.A. (eds.). Lecture Notes in Mathematics,
vol. {\bf 1060.} Berlin Heidelberg, New York: Springer (1984) }

\lref\leclair{LeClair, A.:
Restricted Sine-Gordon theory and the minimal conformal
series. Phys. Lett. {\bf B230}, 103-107 (1989)}

\lref\bax{Baxter, R.J.:
Exactly Solved Models in Statistical Mechanics,
Academic Press, London (1982) }

\lref\Bax{Baxter, R.J.:
Partition function of the eight-vertex lattice model.
Ann. Phys. {\bf 70}, 193--228 (1972) }

\lref\YBE{Jimbo, M. (ed.):
Yang-Baxter equation in integrable systems,
World Scientific, Singapore (1989)}

\lref\BaxSOS{Baxter, R.J.:
Eight-vertex model
in lattice statistics and
one-dimensional anisotropic
Heisenberg chain 1. Some fundamental 
eigenvectors. Ann. Phys. {\bf 76}, 1-24 (1973)\semi
2. Equivalence to
a generalized ice-type model. Ann. Phys. {\bf 76}, 25-47 (1973)\semi
3. Eigenvectors of the transfer matrix and Hamiltonian.
Ann. Phys. {\bf 76}, 48-71 (1973)}

\lref\abf{Andrews, G., Baxter, R. and
Forrester, J.: Eight-vertex SOS model
and generalized Rogers-Ramanujan identities.
J. Stat. Phys. {\bf 35}, 
193-266 (1984)}

\lref\lp{Lukyanov, S.  and   Pugai, Ya.: Bosonization
of ZF algebras: Direction toward deformed Virasoro
algebra.
Rutgers preprint RU-94-41 (1994) (hep-th/9412128)}

\lref\fre{Frenkel, E. and Reshetikhin, N.: Quantum affine algebras
and deformations of the Virasoro and \ $W$-algebras. Preprint (1995)
(q-alg/9505025)}

\lref\yap{Shiraishi, J., Kubo, H., Awata,
H. and Odake, S.: A quantum deformation of the
Virasoro algebra and the
Macdonald symmetric functions.
Preprint YITP/U-95-30, DPSU-95-5,
UT-715 (1995) (q-alg/9507034)}

\lref\ff{
Feigin, B. and Frenkel, E.:
Quantum\ $ W$-algebras and elliptic algebras.
Preprint (1995) (q-alg/9508009)} 

\lref\DotsFat{ Dotsenko,
Vl. S. and Fateev, V. A.: Conformal algebra and
multi-point correlation functions in 2d statistical models. Nucl. Phys.
{\bf B240}
\ [{\bf  FS12}], 312-348 (1984) 
\semi Dotsenko, Vl. S. and  Fateev, V. A.:
Four-point
correlation functions and the operator algebra in 2d conformal invariant
theories with
central charge $c\le1$. Nucl. Phys.
{\bf B251}\ [{\bf FS13}] 691-734
(1985)}

\lref\felder{Felder, G.: BRST approach to minimal models.
Nucl. Phys. {\bf B317}, 215-236 (1989)}

\lref\baxter{Baxter, R.J.: Solvable eight-vertex
model on an arbitrary planar lattice.
Philos. Trans. Roy. Soc.
London Ser. { \bf A289}, 315-346 (1978)}

\lref\rochacaridi{Rocha-Caridi, A.:
Representation theory of the Virasoro
and super Virasoro algebras:
irreducible characters.
In:
Infinite Lie Algebras and Conformal Invariance In
Condensed Matter and Particle Physics,
Proceedings, Bonn, 59-80 (1986)}

\newsec{Introduction}
We start with an extended formulation
of the problem which is attacked by
solving in this work.
\par\noindent
\subsec{General RSOS model}
\par
Knowing the equilibrium structure of the 
crystal-vapor interface is a prerequisite for
understanding and controlling the growth of
crystals from vapor or
dilute solution. At low
temperatures the interface is expected
to be atomistically flat to
minimize the surface energy, whereas at high
temperatures it is 
rough and contains steps producing the  configuration
entropy. This change in the  structure is the
roughening transition of the
interface.
The simple model generally accepted to
depict  the necessary minimum of configurational
degrees of freedom describing the roughening transition
is so-called "Restricted solid-on-solid" (RSOS) model  \ \cristal.
In this model crystal atoms are supposed to be
located at the square lattice sites numerated
by variables \ $a$\  without
allowing an empty space in the
bulk of the crystal. The interface is characterized
by the height or number of crystal layers
$m_a$\ measured from some reference level.
\noindent
The
statistics of the model is governed by 
local positively defined Boltzmann weights
\vskip 0.4cm
\centerline{\epsfxsize 1.5 truecm
\epsfbox{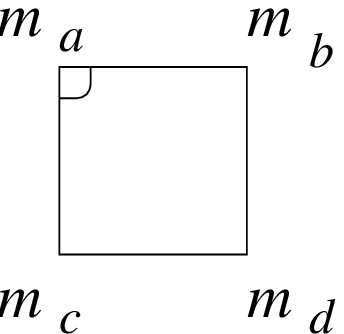}}
\eqn\kiuy{ {\bf W}\left[\matrix{m_a &m_b\cr m_d& m_c} \ \right]=
\exp\left\{-\frac{\varepsilon(m_a,m_b,m_c,m_d)}{k_B T}\right\}\ }
\vskip 0.5cm
\noindent
assigned to every configuration
\ $(m_a,m_b,m_c,m_d)$\ of heights round a face with the
sites \ $(a,b,c,d)$\ 
ordered clockwise from the upper left.
The restriction condition on the weights \ \kiuy\  
expresses that no strongly
fluctuating configurations are allowed. More precisely,
if \ $m$'s\ take integer values between 1 and \ $r-1$\ ,
\eqn\kio{1\leq m_a\leq r-1\ ,}
with the integer \ $r\geq 4$, 
then \ \kiuy\  is non zero only if any two neighboring heights
around the face differ by\ $\pm 1$\ :  
\eqn\miou{|m_a-m_b|=1\ .}
The Boltzmann weight of a  given configuration
is  a  product of weights assigned to
each 
face of the lattice.
At  low temperatures the interface   
tends to take on a height  configuration
minimizing
the total surface  energy
\eqn\hydtd{{\cal E}=\sum_{faces (a,b,c,d)}\varepsilon
(m_a,m_b,m_c,m_d)\ .}
Such configuration is  usually
referred  as a  ground state.
In principle, the system  may leave  a room for 
a number of different ground states
and we
enumerate them  by an  integer  \ $l$.

Consider the following lattice.
\vskip 0.5cm
\centerline{\epsfxsize 3.0 truecm
\epsfbox{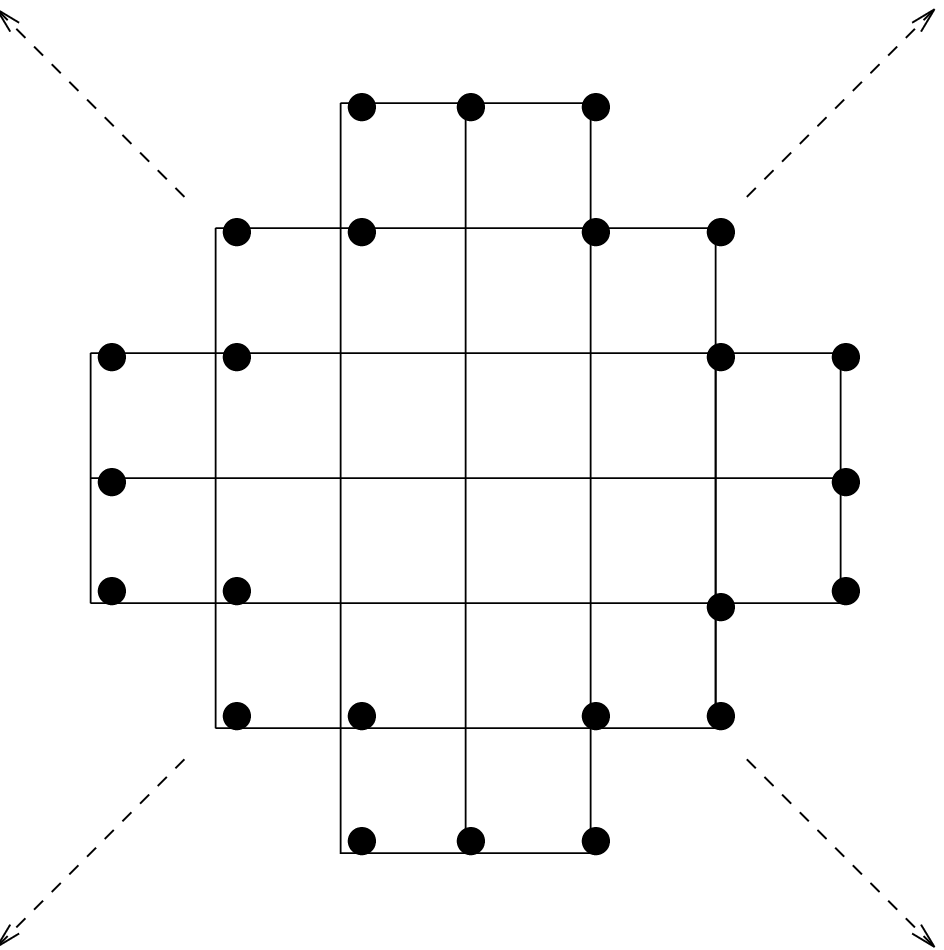}}
\vskip 0.5cm
\noindent
The bullets here denote boundary heights which would be fixed 
somehow. In the present work we assume that the boundary heights 
take the same values as
those in one of the ground states\ \bax .
So, there is one-to-one correspondence between these boundary
conditions and the ground states  and they
can be labeled by the same index\ $l$. 
A partition function for the RSOS model with the 
designated \ $l$-th boundary condition  
is given by 
\eqn\part{Z^{(N)}_l=\sum_{heights\{m\}}\ \prod_{faces(a,b,c,d)} 
{\bf W}\left[\matrix{m_a &m_b\cr m_d& m_c} \ \right]\ .}
Here \ $N$\ is a number of the lattice faces. 
Of particular interest is
the partition function\ $Z^{(N)}_l$\ 
at the  thermodynamic limit
\  $N\to \infty$,
when
the lattice uniformly increases
in all directions.
The short
distance interaction ensures, that 
\eqn\loi{Z^{(N)}_l\to  \kappa^N Z_l\ ,\ \ \ N\to\infty\ ,}
and\ $ \kappa$ \  is
a partition function per face. The above mentioned
boundary conditions  break explicitly
the translation invariance, hence 
$Z_l$ \ in \loi  \ may depend
on them,
in contrary to\ $\kappa $
\foot{
The translation invariance can be restored
by a proper  averaging   over
all possible values  of \ $l$\ .}.

Other important   quantities  for
studying
are\ $n$-point Local Height Probabilities (LHP).
They can be defined as follows; Let 
us specify \ $n$\ reference sites
belonging the same vertical column
and
label them from down to up  by\ $ 1,2,...n$.
The probability that heights at
these sites
have  the values
\ $1\leq k_1,... k_n\leq r-1$\ 
reads:
\eqn\cf{
P^{(N)}_{k_1,... k_n}(l)=
\Big[Z^{(N)}_l\Big]^{-1}\sum_{heights \{m\}}\
\prod_{j=1}^{n}\delta(k_j,m_j)\ \prod_{ faces(a,b,c,d)}
{\bf W}
\left[\matrix{m_a&m_b\cr m_d&m_c}\right]\  .\ }
Again,
of prime importance  are
LHP  at
the thermodynamic limit
\eqn\mcnvb{P_{k_1,... k_n}(l)=
\lim_{N\to\infty}P^{(N)}_{k_1,...k_n}(l)\ .}

\subsec{ABF  model}
The Boltzmann
weights\ \kiuy\
are 
free parameters for the RSOS model,
and an exact analytical calculation of
LHP \ \mcnvb\  for their general values 
does not seem to be a  realizable
problem nowadays. 
Nevertheless, there are some 
special cases 
when the  model is a solvable one.
As it  usually is, a  local integrability  condition
is given by
the  set of algebraic  equations
for the  Boltzmann weights \bax, \YBE 
\eqn\ybe{\eqalign{
&\sum_{m}  {\bf W}\left[\matrix{m_1& m\cr m_2& m_3}\biggl|\ u-u'
\ \right]
{\bf W}\left[\matrix{m_1 & m_4\cr m &m_5}\biggl|\ u'
\ \right]
{\bf W}\left[\matrix{m & m_5\cr m_3 & m_6}\biggl|\ 
u
\ \right]=\cr
&\sum_{m}{\bf W}\left[\matrix{m_1 & m_4\cr m_2 & m}\biggl|\ 
u
\ \right]
{\bf W}\left[\matrix{m_2 & m\cr m_3 & m_6}\biggl|\ 
u'
\ \right]
{\bf W}\left[\matrix{m_4 & m_5\cr m & m_6}\biggl|\
u-u'
\ \right]\ .
}}
\vskip 0.4cm
\centerline{
\epsfxsize 8.0 truecm
\epsfbox{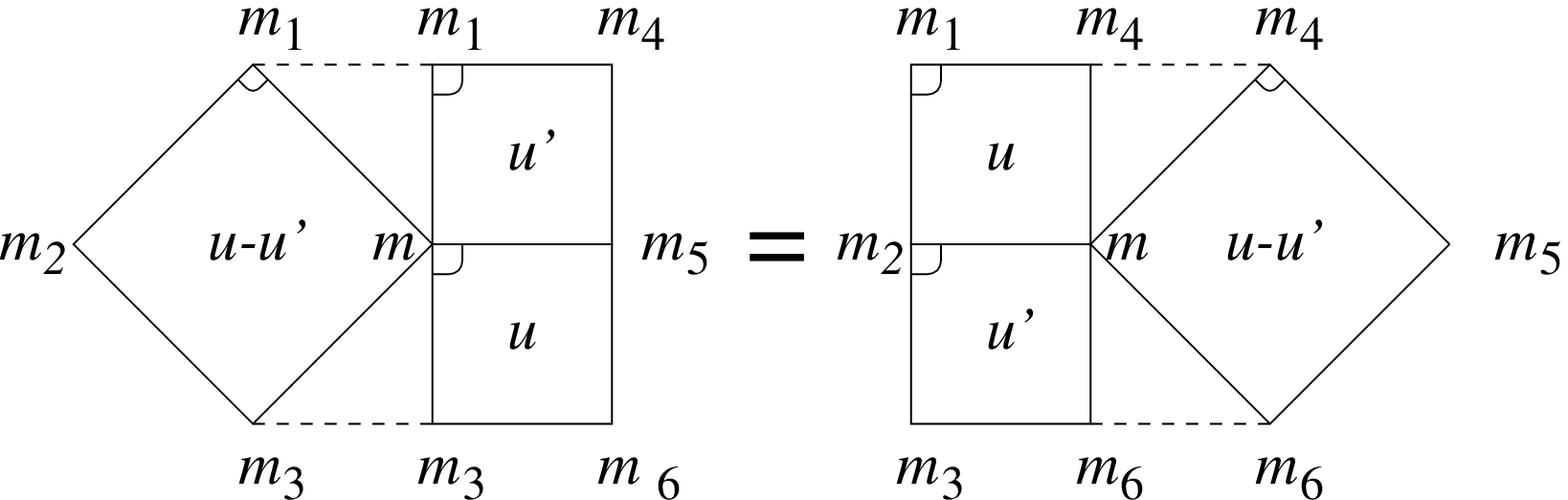}
}
\vskip 0.5cm
\par\noindent
Here \ $u$\ is some
variable
parameterizing  the  Boltzmann weights.
Andrews\ {\it  et al.} \refs{\abf} have succeeded in  finding
of   solutions of the Yang-Baxter equation (YBE)\  
\ybe\  for any integer \ $r\geq 4$.
In addition to the spectral parameter\ 
$u$,  the ABF solution depends
on a parameter \ $p$.
It gives a two parametric family in a  manifold of the 
Boltzmann  weights of the RSOS model.
The  integrable  weights read explicitly:
\eqn\iu{\eqalign{
&{\bf W}
\left[\matrix{m\pm 2&m\pm 1\cr m\pm1  &m}
\right]=R\ ,\cr
&{\bf W}
\left[\matrix{m&m\pm 1\cr m\pm1  &m}
\right]=R\ \frac{[m\pm u]}{[1-u]
[m]}\ ,\cr
&{\bf W}
\left[\matrix{m&m\pm 1\cr m\mp1  &m}
\right]=R\ \frac{\big([m+1][m-1]\big)^{\frac{1}{2}}}{[m]}\ 
\frac{[ u]}{[1-u]}
\ .}}
Here we use the  short notation\ 
\eqn\oiuy{[u]=h(u)/h(1)\ ,}
with
$$h(u)=2\  |p|^{1/8}
\ \sin(\pi u/r) \prod_{n=1}^{+\infty} \big(1- p^{n}\big)
\ \big(1- 2 p^{n} \cos(2\pi u/r)+p^{2n}\big)\ .$$
In order to provide the convergence
of the  infinite product,
the parameter \ $p$\ has to be restricted to the domain
$$-1<p<1 \ . $$
The positive factor \ $R$\ in\  \iu\ determines
a normalization of the weights.

The ABF
weights family consists of
two distinct  manifolds
with\ $0<u< 1$, \ and $-1<u<0$. 
According to \refs{\abf}, the  manifolds 
are  divided also into   phases (regimes)
by the line of critical points \ $p=0$.
In this work we restrict our attention to the 
so-called regime III:
\eqn\hsyt{ 0<p<1\ ,\ \ \  0<u< 1\ .
 } 
The parameter \ $p$\ 
measures a  deviation from the criticality
\foot{In the case of the regime III,
the scaling  behavior in the vicinity
of the critical line is described by 
the Quantum Field Theory and referred as
the Restricted Sine-Gordon model  
\ \lref{\zamolo} ,
\leclair .}.
The value \ $p=1$\
can be associated   with the zero temperature,
since in this case  the system
is frozen in one 
of the ground states.
As it follows from
\iu ,
each ground state
in the regime III  has  the form:
\vskip 0.5cm
\centerline{
\epsfxsize 4.0 truecm
\epsfbox{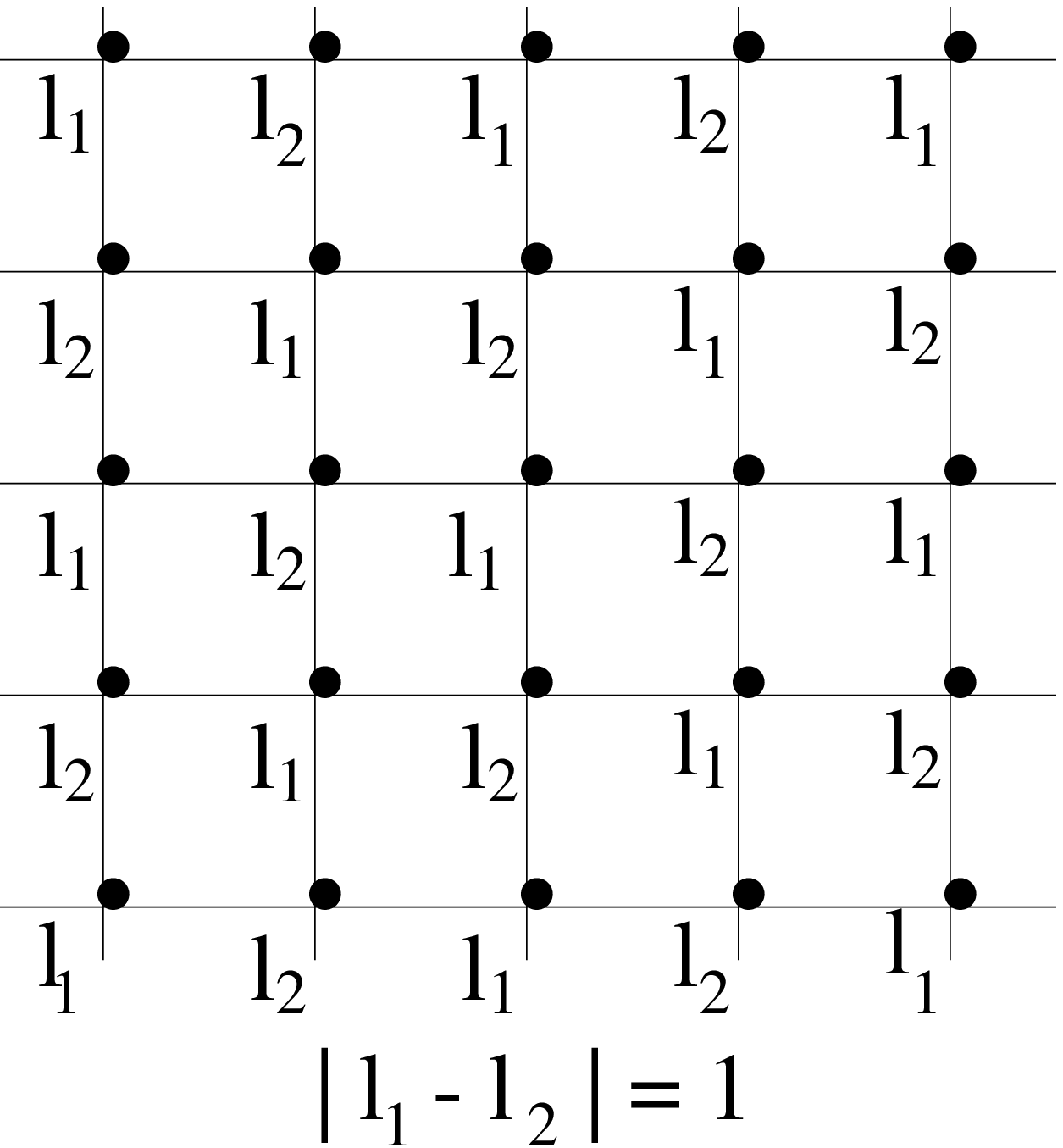}
}
\vskip 0.5cm
\par\noindent
Therefore, we identify the integer number\ $l$,   
enumerating the ground states,
with\ $min(l_1,l_2)$ \  and
\eqn\hdyft{1\leq l \leq r-2\ .}
The same number specifies
also the boundary conditions,
used in the definition of  LHP\ \cf ,\   \mcnvb.

Now, let us break down the \ $n-1$\ rows between the reference
points \ $ a=1,2,...,n$ and replace the spectral parameter
\ $u$\  in
these rows to \ $0<u_1,...,u_{n-1}<1 $,  respectively.
\vskip 0.5cm
\centerline{
\epsfxsize 4.0 truecm
\epsfbox{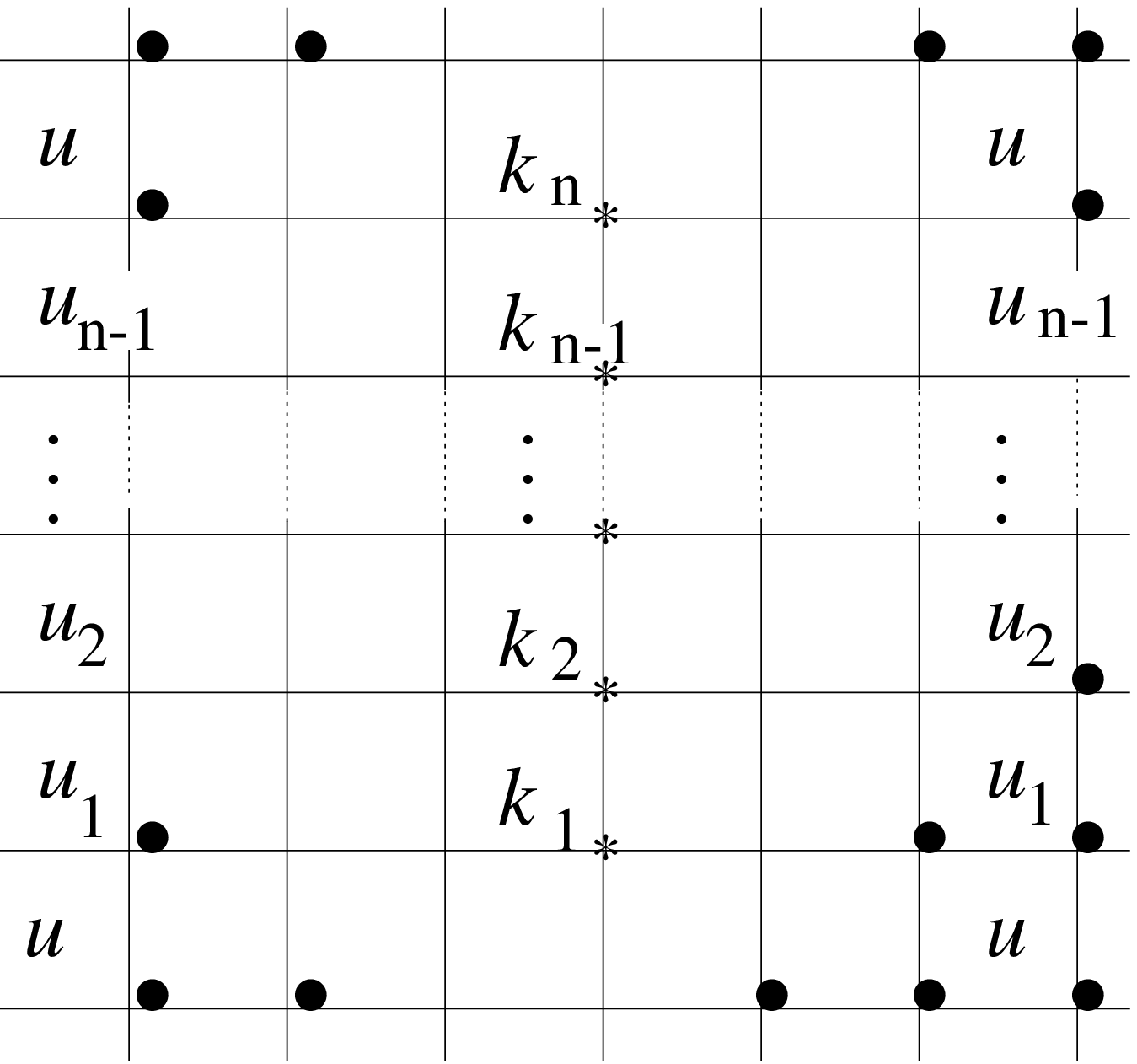}
}
\vskip 0.5cm
\par\noindent
As well as  in the  homogeneous case, one can introduce
$n$-point
LHP $P_{k_1,...k_n}(u_1,...,u_{n-1}|l)$
at the thermodynamic limit
on the lattice
with   such kind of dislocations. In this work we are 
going to study these quantities. 
Our main result 
is an
integral representation
for them
given by the formulas (5.11).
Notice, that
it would be  convenient for us to  use 
the parameters \ $\zeta$ and $x$
\eqn\jdufyt{  x =\exp\left[\frac{2\pi^2}{r\ln p}
\right]\ ,\ \  \zeta=x^{2u}\ ,  }
rather then
 \ $u$\ and\ $p$.
Hence,  the  \ $n$-point
LHP
for the inhomogeneous
lattice 
are denoted as
$$P_{k_1,...k_n}(\zeta_1,...,\zeta_{n-1}|l)\ $$
in what
follows. It is worth to point out 
here,  that due to  YBE\ \ybe\  these functions depend 
on the ratios \ $\zeta_j\zeta^{-1}_1$\ only \refs{\baxter}.
\newsec{Corner Transfer Matrix and one-point LHP}
\subsec{Corner Transfer Matrix}
The key calculation tool  for 
LHP is the Corner Transfer Matrix (CTM) \Baxeight ,
\bax , \abf .
In this section we concentrate on finding  the one-point
function
to give an idea on this important notion.  
Consider the square  lattice
with the  reference site \ $ O$\  as being the central one.
\vskip 0.5cm
\centerline{
\epsfxsize 4.0 truecm
\epsfbox{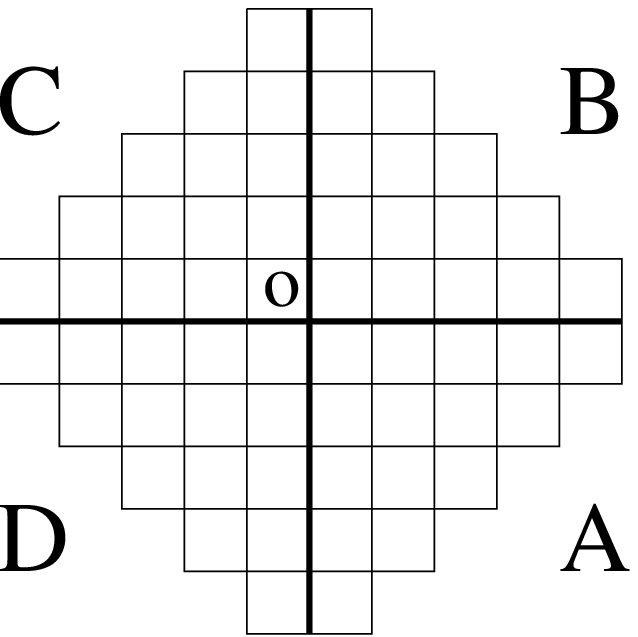}
}
\vskip 0.5cm
\par\noindent
The probability \ $P_k(l)$\ 
that the  central height
equals to \ $k$\ is determined by
\  \cf , \ \mcnvb\   with\ $ n=1$. 
One can divide the whole lattice into
four quadrants by the horizontal and  vertical lines
intersecting  at\ $O$.
The idea is to calculate the sum in
\ \cf \ in two steps. First, let us 
perform summation over the  sites
inside the quadrants with given values
of heights along the cuts. 
Doing so, one  treats
the resulting expressions as matrix elements of 
the four Corner Transfer Matrices \ $ \bf A,B,C,D$\ 
associated with the corresponding 
quadrants. The second step is to find the sum of heights 
on the boundaries between the quadrants.
More explicitly, 
consider the lower right quadrant with the hight configuration
$${\bf m}=(m_1,m_2,...l_1,l_2)\ ,\ \ \ m_1=k$$ 
along the horizontal 
boundary 
and
$${\bf m}'=(m'_1,m'_2,...l_1,l_2)\ ,\ \ \ m'_1=k$$
for the vertical one.
\vskip 0.5cm
\centerline{
\epsfxsize 5.0 truecm
\epsfbox{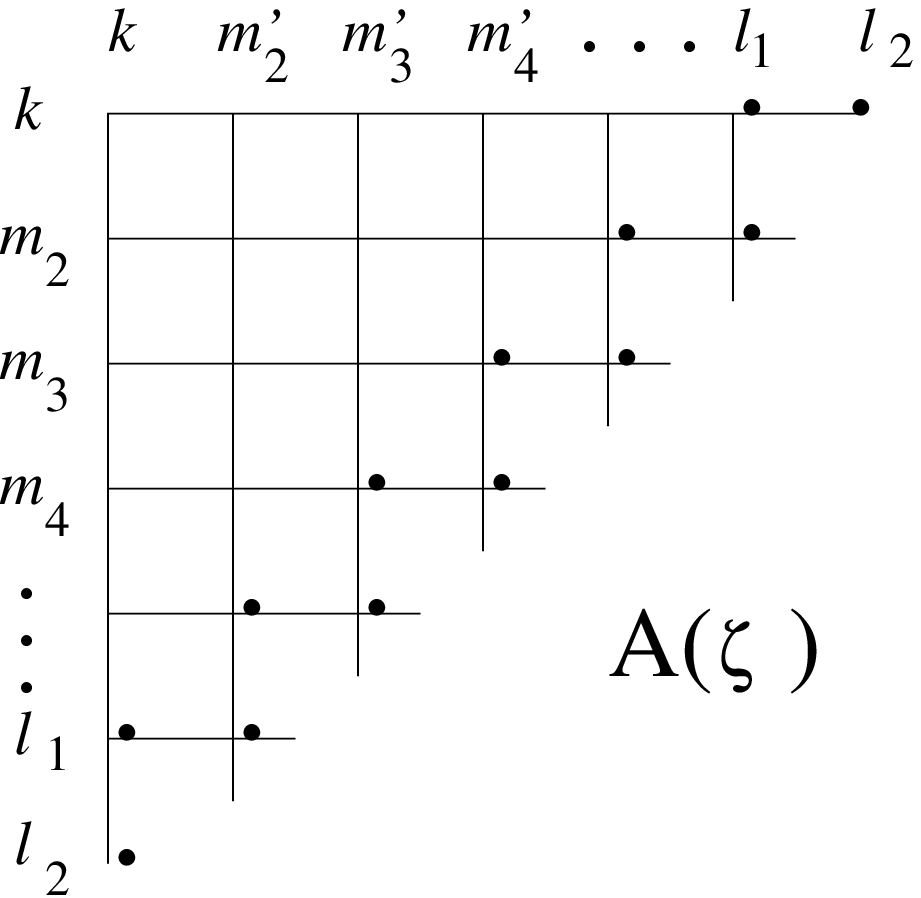}
}
\vskip 0.5cm
\par\noindent
The
Corner Transfer Matrix 
\ $\bf{A}(\zeta)$\ is a matrix
with
elements
\ $\big[{\bf{A}}\big]_{{\bf m}}^{{\bf m}'}$\ given
by the 
partition functions of the quadrant with the  fixed boundary heights.
Notice, that
we  specify the dependence
on the  spectral parameter chosen in the form
\ \jdufyt .
CTMs \ 
${\bf{B}}(\zeta), {\bf{C}}(\zeta), {\bf{D}}(\zeta)
$\
for other quadrants are defined similarly.
By the definition,
CTM 
breaks up into \ $r-1$\ diagonal blocks 
with the given value of 
the
central height. Let \ ${\bf P}_{k}$\ be the 
diagonal matrix which 
diagonal entries are units for the block with \ $m_1=k$\ 
and all other elements being zero. Then the remaining sum 
in \ \cf \ over the heights at the cuts 
consists in  taking the trace:
\eqn\ctm{
P_k(l)=Tr\Big[\ {\bf P}_{k}\ {\bf DCBA}\ \Big]\Big/
Tr\Big[\ \bf {DCBA}\  \Big]\ . 
}
This expression does not 
depend on a normalization of CTMs. 

According to \Baxeight ,\bax ,
CTMs and their products would be  well defined operators
at the thermodynamic
limit,  if the Boltzmann weights are normalized 
in such a way 
that  the partition function  per face\ $\kappa$\ \loi\ 
equals to one. For the  ABF weights \iu \ in the regime III 
this normalization
means
\refs{\Bax}, \refs{\abf}: 
\eqn\jdut{
R=\zeta^{\frac{r-1}{2r}}\
\frac{
(x^2 \zeta^{-1};x^{2r},x^4)_{\infty} \
(x^{2+2r} \zeta^{-1};x^{2r},x^4)_{\infty}\
(x^4 \zeta;x^{2r},x^4)_{\infty} \
(x^{2r}\zeta;x^{2r},x^4)_{\infty}}
{(x^4 \zeta^{-1};x^{2r},x^4)_{\infty} \
(x^{2r} \zeta^{-1};x^{2r},x^4)_{\infty}\
(x^2 \zeta;x^{2r},x^4)_{\infty} \
(x^{2+2r} \zeta;x^{2r},x^4)_{\infty} }\  ,}
where  the notation 
$$
(\zeta;q_1,q_2,...,q_n)_{\infty}=
\prod_{\{k_i\}=0}^{\infty}(1-\zeta\  q_1^{k_1}q_2^{k_2}...q_n^{k_n})
$$
is used.
The function
\ $R(\zeta)$ \ obeys the  set of relations
\eqn\ksiudy{\eqalign{&R(\zeta)R(\zeta^{-1})=1\ ,\cr
&R(x^2\zeta^{-1})=\big(x\zeta^{-1}
\big)^{\frac{r-1}{r}}\ 
\frac{(\zeta;x^{2r})_{\infty}\ (x^{2r}\zeta^{-1};x^{2r})_{\infty}}
{(x^2\zeta^{-1};x^{2r})_{\infty}\ (x^{2r-2}
\zeta;x^{2r})_{\infty}}\ R(\zeta)\ .}}
With this normalization,
the ABF weights
satisfy 

{\it - unitarity relation}
\eqn\hdfa{\sum_{m_2}\ 
{\bf W}\left[\matrix{m_1&m_2\cr m_4  &m_3}
\biggl|z   \right]\ 
{\bf W}\left[\matrix{m_1&m'_4\cr m_2  &m_3}
\biggl|z^{-1}   \right]=\delta_{m_4, m'_4}\ , }

{\it -  crossing symmetry relation}
\eqn\hdfc{{\bf W}
\left[\matrix{m_1&m_2\cr m_4  &m_3}
\biggl|x^2\zeta^{-1}   \right]=
\sqrt{\frac{[m_2][m_4]}{[m_1][m_3]}}
\ {\bf W}
\left[\matrix{m_4&m_1\cr m_3  &m_2}
\biggl|\zeta   \right]\ ,}

{\it - "initial" condition} 
\eqn\hytr{
{\bf W}\left[\matrix{k_1&k_2\cr k_4&k_3}
\biggl|\zeta=1 \right]=\delta_{k_2 k_4}\ .}

{}From this point  on we assume 
the normalization \ \jdut\ holds
and   at the thermodynamic limit 
CTMs are  well defined linear operators
acting in the space covered by vectors
\ $(k,m_2,m_3,...l_1,l_2,l_1,...)$. We denote such
spaces as \ ${\cal L}_{l,k}$:
\eqn\kiyf{{\bf A}(\zeta):\ {\cal L}_{l,k}\to {\cal L}_{l,k}\ . }

\subsec{Spectrum of CTM in   ABF model}
At the thermodynamic limit the Corner Transfer Matrix
for solvable  statistical models  has the  remarkably
simple
form \Baxeight ,\bax
\eqn\thermo{
{\bf A}(\zeta) = \zeta^{{\bf H}_C}\ ,}
where the Corner Hamiltonian \ ${\bf H}_C$\
is a \ $\zeta$-independent operator.
Notice, that
this is a
common property,  since 
it follows from  
YBE\ 
\ybe \ and the crossing symmetry  condition for the Boltzmann
weights.
The relation\ \hdfc
\ allows one
to express the CTMs \ ${\bf B,C,D}$\
in terms of \ ${\bf A}$:
\eqn\cds{{\bf B}(\zeta)=
{\bf S }_{l,k} \ {\bf  A}(x^2 \zeta^{-1})\ , \ \ {\bf C}(\zeta)=
{\bf A }(\zeta)\ ,\ \ 
{\bf D}(\zeta) = {\bf S }_{l,k} \ {\bf A}(x^2 \zeta^{-1})\  .}
The matrix \ ${\bf S}_{l,k}$ depends
on the  value of
the central height\ $k$\  and the boundary condition only:
\eqn\kio{ {\bf S }_{l,k}=\sqrt{[k] {\hat [}l{\hat ]}}\  {\bf I}\ .}
Here \ ${\bf I}$\ is the  identity  matrix.
An explicit form of the
function \ ${\hat [}l{ \hat ]}$\ is inessential
for the problem under consideration, since it is cancelled 
out  
in the final expressions.

Using the formulas \ \thermo-\kio ,   we can rewrite \ctm
\ as:
\eqn\kuytb{P_k(l)=[k]\ Tr_{{\cal L}_{l,k}}\Big[\  x^{4{\bf H}_C}
\ \Big]\Big/\biggl\{\sum_{m=1}^{r-1}\ [m]\ 
Tr_{{\cal L}_{l,m}}\Big[\  x^{4{\bf H}_C}
\ \Big]\biggr\}\ .}
Therefore, the 
calculation is reduced to finding the
spectrum of  the
Corner Hamiltonian \ ${\bf H}_C$
in
the space\ ${\cal L}_{l,k}$.
At the
thermodynamic limit the spectrum is
notable  
for the following reasons;
It is
\eqn\spe{\eqalign{
&
\ -\ bounded \ from\ \ below\ ,\cr 
&\ -\ discrete\ ,\cr 
&-\ equidistant\ .}}
\noindent
These features seem to be rather general ones for
solvable lattice models \refs{\bax},\ \YBE
\foot{It is a
starting point for so-called Angular Quantization approach
for the  integrable Quantum Field Theory\ \refs{\singordon}.
In the scaling limit
the Corner Hamiltonian coincides
(up to a  multiplicative
normalization)   with the
Lorentz boost generator,
acting in the Angular Quantization space.}.
In the ABF case,
they follow from the 
fact that the Boltzmann weights 
are quasiperiodic functions
under the changing 
\  
$\zeta\to e^{2\pi i} \zeta\ .$
Andrews {\it et al.}\  \refs{\abf}\ 
have shown
that the 
eigenvalues of the Corner  Hamiltonian \ ${\bf H}_C$\
in the regime III  have the form:
\eqn\juyt{\Delta_{l,k}-\frac{c}{24}+m\ , \ \ m=0,1,2,...\ ,}
where
\eqn\loiuy{\Delta_{l,k}=\frac{\big(rl-(r-1)k\big)^2-1}
{4r(r-1)}\ , 
\ c=1-\frac{6}{(r-1)r}\ .}
The discreteness of the spectrum implies that it
does not change discontinuously with the temperature parameter
\ $p$. Hence, to find the  multiplicities
of  eigenvalues \ \juyt\
of the Corner Hamiltonian acting in the space
\ ${\cal L}_{l,k}$\  \juyt\   
it is enough to 
perform the computation
for any value of \ $p$\ \hsyt .
They  can be  derived at  low temperatures
limit \ $p\rightarrow 1$\
by combinatorial methods \abf .
The amazing observation \huse  ,\ \refs{\japAn},\ \refs{\saleur}
is  the spectrum of \ ${\bf H}_C$\  
{\it coincides}\ with 
one of the Virasoro algebra generator \ $L_0-\frac{c}{24}$\ 
acting in the irreducible representation
with the  highest weight\ $\Delta_{l,k}$\  and the
central charge\ $c$\  
\loiuy . This  immediately leads to the 
formula for the one-point LHP\ \abf 
\eqn\ouy{P_k(l)={\bar Z}^{-1}_l\
[k] \  \chi_{l,k}(x^4)\ , }
where\foot{Notice, that \ $Z_l={\hat [}
l{\hat ]}\  {\bar Z}_l$, where the function
$ Z_l$\ is  defined by\ \loi\  and 
\ ${\hat [} l\hat{ ]}$\ 
is the same as in\ \kio .  }
\eqn\zl{
{\bar Z}_l=\sum_{m=1}^{r-1}\  [m]\ \chi_{l,m}(x^4)\ .
}
Here
we denote
the  character of the  irreducible module 
\refs{\rochacaridi} by
\eqn\char{\eqalign{
\chi_{l,k}(q)=
(q;q)^{-1}_{\infty}\ q^{-\frac{c}{24}}\ 
\Big\{ &q^{\Delta_{l,k}}
E(- 
q^{rl-(r-1)k+(r-1) r}, q^{2 (r-1) r})-\cr &
q^{\Delta_{-l,k}}
E(- 
q^{-rl+(r-1)k+(r-1)r}, q^{2 (r-1) r})\Big\}\ .}}
and use the notation for the  elliptic function 
\eqn\trassa{
E(z,q)=(q;q)_{\infty}(z;q)_{\infty}(qz^{-1};q)_{\infty}\ .}

\newsec{Vertex operators and multi-point 
Local Height Probabilities}
\subsec{Vertex operators } 
Now we begin to study the multi-point
LHP. In this subsection we follow
the ideas of the work\ \japising . Consider
the inhomogeneous lattice.
Our challenge is to find the probabilities that heights
in \ $n$\ successive sites from the same vertical column
take values \ $k_1,\dots  k_n$. It is useful to divide 
the lattice into \ $2n+2$\  parts, as it is shown in the picture:
\vskip 0.5cm
\centerline{
\epsfxsize 10.0 truecm
\epsfbox{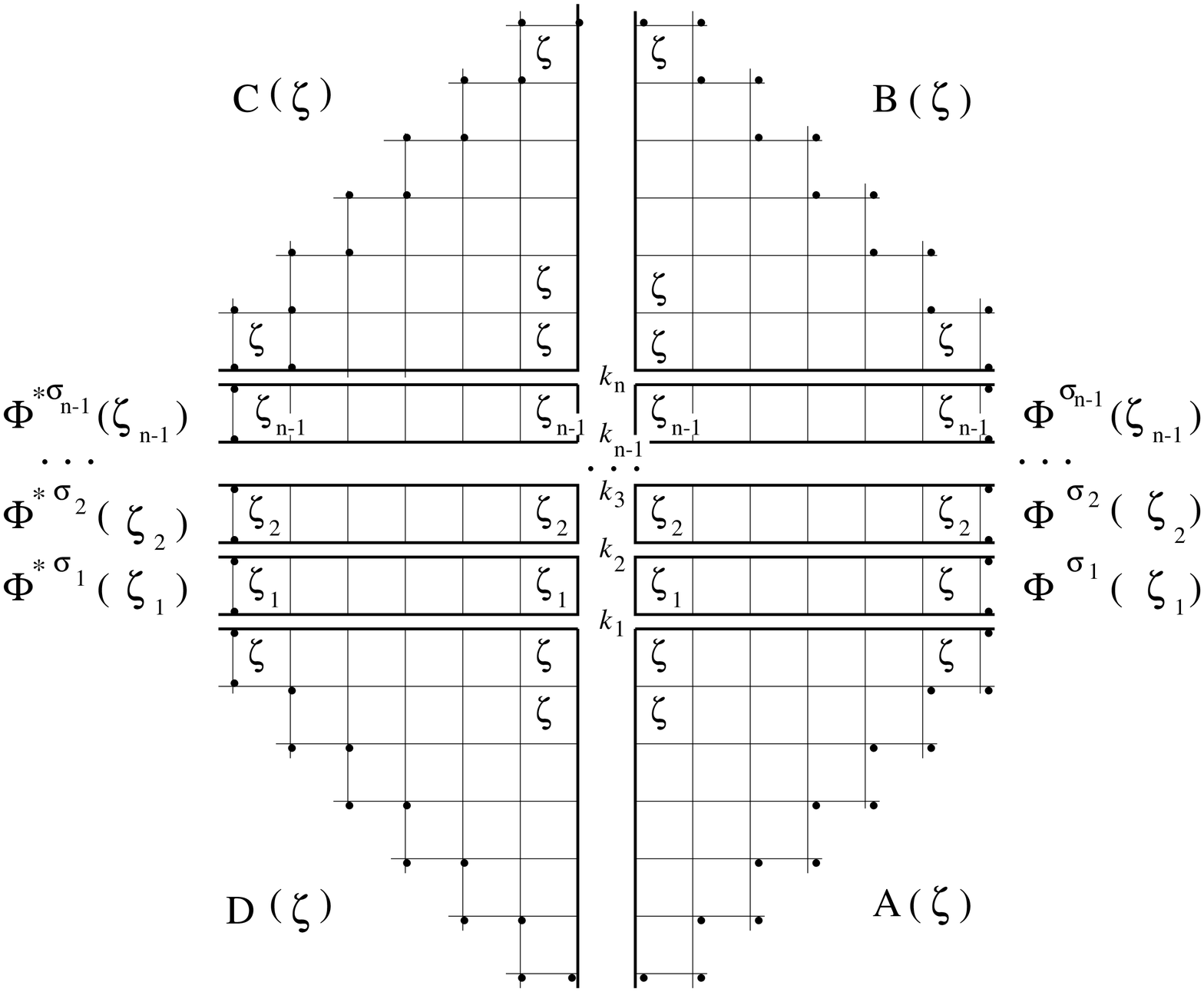}}
\vskip 0.5cm
\par\noindent
Again, at the first step of the calculation we specify the
heights along the cuts and
introduce  the  partition functions  for
the corresponding parts. 
The four of them are CTMs. 
Others
are given by the  products of the 
weights
in the form of half infinite lanes:
\vskip 0.5cm
\centerline{
\epsfxsize 8.0 truecm
\epsfbox{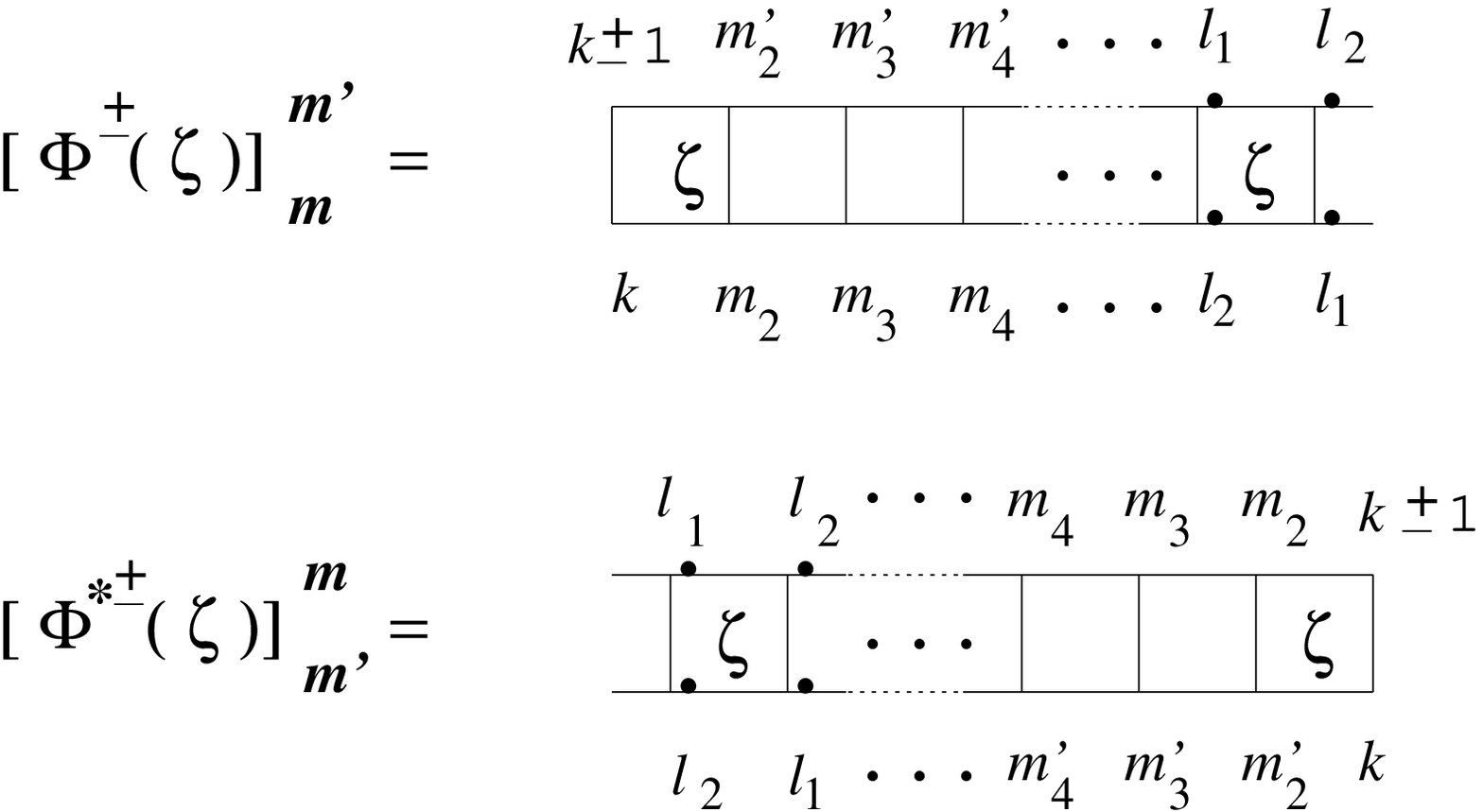}}
\par\noindent
We treat such half infinite 
products
with a given values of
the spectral parameter\ $\zeta$,
as  
matrix elements
of operators 
\ ${\bf \Phi}^{\pm}(\zeta)$\ ,
\ ${\bf \Phi}^{*\pm }(\zeta)$. The operators
intertwine the spaces 
\ ${\cal L}_{l,k}$\ and \ ${\cal L}_{l,k\pm 1}$
\eqn\inter{\eqalign{
&{\bf \Phi}^{\pm}(\zeta):\  {\cal{L}}_{l,k}\rightarrow 
{\cal{L}}_{l,k\pm 1}\cr
&{\bf \Phi}^{*\pm }(\zeta):\  {\cal{L}}_{l,k\pm 1}\rightarrow 
{\cal{L}}_{l,k}
\ \ ,} }
and are referred as
vertex operators (VO).
The second step is to carry out the  sum of heights over the cuts
assuming that variables at the sites \ $1,...n$\
equal to \ $k_1,...k_n$\ , respectively. 
As before, 
it is equivalent to taking a trace.
In such a way,
we obtain the expression for the multi-point
LHP: 
\eqn\mnhy{\eqalign{
&P_{k _1, ...k_n}(\zeta_1,...\zeta_{n-1}\  |l)
=\biggl\{\sum_{m=1}^{r-1}\ Tr_{{\cal{L}}_{l,m}}\Big[\ 
{\bf D}(\zeta){\bf C}(\zeta){\bf B}(\zeta){\bf A}(\zeta)\ \Big]
\biggr\}^{-1}\times\cr
&Tr_{{\cal{L}}_{l,k_1}}\Big[\ {\bf D}(\zeta)
{\bf \Phi}^{*\sigma_1}(\zeta_1)
..{\bf \Phi}^{*\sigma_{n-1}}(\zeta_{n-1}){\bf C}(\zeta)
{\bf B}(\zeta){\bf \Phi}^{\sigma_{n-1}}(\zeta_{n-1})	
..{\bf \Phi}^{\sigma_1}(\zeta_1){\bf A}(\zeta)\ \Big]\ ,}}
where\ $\sigma_{s}=k_{s+1}-k_{s}$.

To proceed further, one need to study the properties of VO.
They
can be derived 
from the heuristic graphic arguments\  \refs{\japising}.

\noindent
{\it I. Commutation relations}

\noindent
This property  follows  from the
Yang-Baxter equation\ \ybe . 
\vskip 0.5cm
\centerline{
\epsfxsize 10.0 truecm
\epsfbox{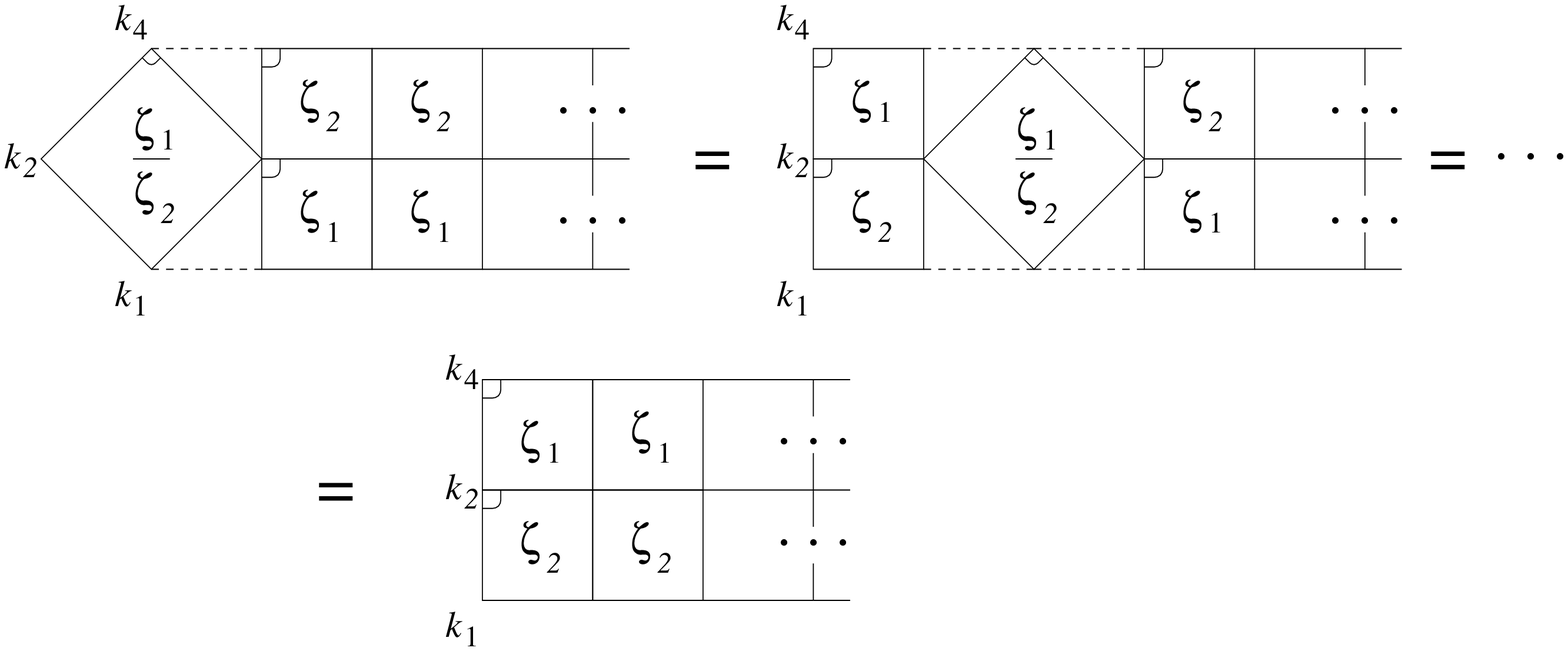}
}
\vskip 0.5cm
\par\noindent
The face
added in the first
figure is pushed up to the infinity due to  YBE.
At the operator notations this implies 
that \ ${\bf \Phi}^\pm(\zeta)$\  
constitute the  Zamolodchikov-Faddeev
algebra in the Interaction Round a Face  form: 
\eqn\zfal{\eqalign{
{\bf \Phi}^{k_4-k_2}(\zeta_1)&
{\bf \Phi}^{k_2-k_1}(\zeta_2)|_{{\cal{L}}_{l,k_1}}=\cr
&=\sum_{k_3}{\bf W}
\left[\matrix{k_4&k_3\cr k_2&k_1}\biggl|\  \zeta_1\zeta^{-1}_2
\ \right]\ 
{\bf \Phi}^{k_4-k_3}(\zeta_2){\bf \Phi}^{k_3-k_1}(\zeta_1)|_
{{\cal{L}}_{l,k_1}}\ .}}

\noindent
{\it II. Homogeneity  condition }

\noindent
Consider the
composition of CTM and VO:  
\vskip 0.5cm
\centerline{
\epsfxsize 7.0 truecm
\epsfbox{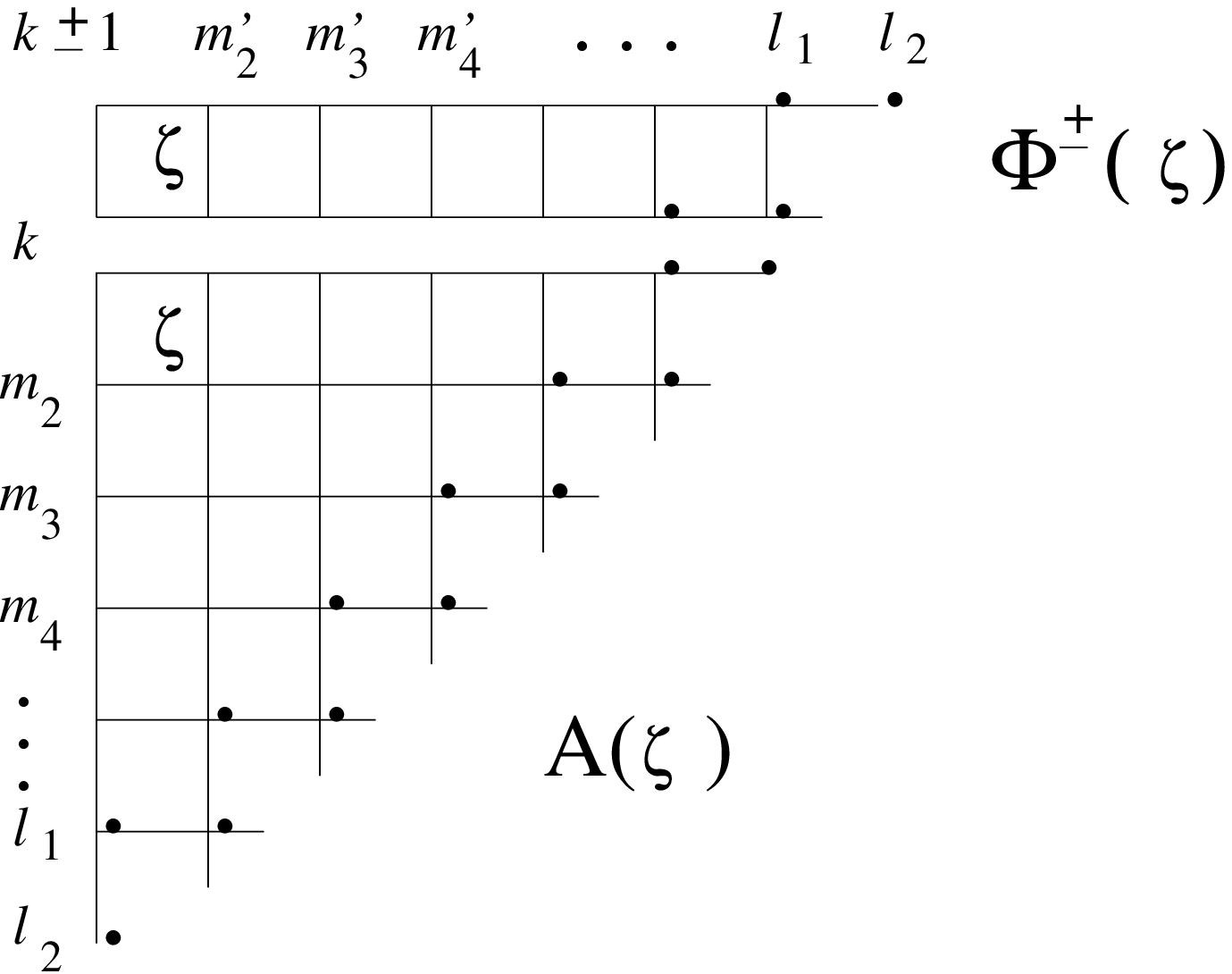}
}
\vskip 0.5cm
\par\noindent
At the  thermodynamic limit this would be CTM
\ ${\bf A}(\zeta)$\ defined, however, in the space 
\ ${\cal{L}}_{l,k\pm 1}$\ rather
than in \ ${\cal{L}}_{l,k}$\ .
To obtain an equality,
these spaces must be intertwined.
It is provided by
action of the operator \ ${\bf \Phi}^{\pm}(1)$\ 
on the space \ ${\cal{L}}_{l,k}$,
because of the "initial" condition\ \hytr .
Therefore,
we have
$${\bf \Phi}^{\pm}(\zeta)\ 
{\bf A}(\zeta)|_{{\cal{L}}_{l,k}}
={\bf A}(\zeta)\ {\bf \Phi}^{\pm}(1)|_{{\cal{L}}_{l,k}}\ . $$
This leads to the equation:
\eqn\der{
z^{-{\bf H}_C } {\bf \Phi}^{\pm}(\zeta)\ z^{{\bf H}_C}
={\bf \Phi}^{\pm}(\zeta z^{-1})\ .}

\noindent
{\it III. Normalization condition}

\noindent
To obtain the last property of VO we need to use 
the unitarity \ \hdfa\  and
crossing symmetry\ \hdfc\  relations. They allow one to get
so-called
normalization condition: 
\eqn\uytiii{\biggl\{\sqrt{[k+1]}\
{\bf \Phi}^-(x^{-2}\zeta){\bf \Phi}^+(\zeta)+
\sqrt{[k-1]}\ {\bf \Phi}^+(x^{-2}\zeta)
{\bf \Phi}^-(\zeta)\biggr\}\Big|_{{\cal L}_{l,k}}=
\sqrt{[k]}\ { \bf I}\  \Big|_{{\cal L}_{l,k}}\   .}

Now, let us express
CTMs
\ ${\bf  B,C,D}$\ and VO \ ${\bf \Phi}^{*\pm}$\
via 
\ ${\bf A}$\ and \ ${\bf \Phi}^{\pm}$, respectively,
by the  formulas\ \cds\  and
\eqn\ckro{
{\bf \Phi}^{*\pm}(\zeta)
= {\bf \Phi}^{\mp}(\zeta)\ .}
The latter follows from the crossing symmetry condition\ \hdfc .
Taking \der \ into account, 
it is easy to find that  
LHPs can be  written in the form\  \japising:
\eqn\hydtr{\eqalign{P_{k_1,...k_n}&(\zeta_1,...\zeta_{n-1}\ |l)
={\bar Z}^{-1}_l\ \sqrt{ [k_1][k_n]}\times\ 
\cr
Tr&_{{\cal L}_{l,k_1}}\Big[\ x^{4 {\bf H_C}}\ 
{\bf \Phi}^{-\sigma_1}(x^{-2}\zeta_1)...
{\bf \Phi}^{-\sigma_{n-1}}(x^{-2}\zeta_{n-1})\ 
{\bf \Phi}^{\sigma_{n-1}}(\zeta_{n-1})...
{\bf \Phi}^{\sigma_1}(\zeta_1)\ \Big]\ ,}}
where\ $\sigma_s=k_{s+1}-k_{s}$,
and \ ${\bar Z}_{l}$ is given by \zl . 
\subsec{ Two-point LHP}
The simplest example of LHP \hydtr\ 
with \ $n=1$\
was treated in the previous section. Now  
we would like to show that from the properties of VO
and CTM 
we can deduce the two-point function.
As well as the  one-point function, the
two-point LHP do not depend 
on the spectral parameter.  
Then, according to \  \hydtr ,   
the probabilities that heights
at two adjacent sites 1 and 2 
are equal to \ $k, k\pm 1$,  respectively, are given by:
\eqn\uytr{P_{k,k\pm 1}(l)= {\bar Z}^{-1}_l\ \sqrt{ [k][k\pm 1]}\ 
Tr_{{\cal L}_{l,k}}\Big[\ x^{4 {\bf H_C}}\
{\bf \Phi}^{\mp}(x^{-2}\zeta)\ {\bf \Phi}^{\pm}(\zeta)\ \Big]\ .}
It follows immediately from \uytiii\ that
these functions obey the equation
\foot{Note, that  (3.9)  
obviously follows from the definition of
the one-point function and \ouy.} 
\eqn\swa{P_{k,k+1}(l)+P_{k,k-1}(l)={\bar Z}^{-1}_l\  [k]
\ \chi_{l,k}(x^4)\ .}
In addition, the  two-point LHP  has to
satisfy the evident requirements:
\eqn\oytr{P_{k,k-1}(l)=P_{k-1,k}(l)\ ,}
\eqn\mjhg{P_{1,0}(l)=0\ .}
The formulae \ \swa\ and\  \oytr \ specify the
two-point functions uniquely:
\eqn\mssy{P_{k+1,k}(l)=
P_{k,k+1}(l)={\bar Z}_{l}^{-1} \sum_{m=1}^{k}(-1)^{k-m}[m]
\ 
\chi_{l,m}(x^4)\ . }
This equation
will lend support to the 
validity of our general 
result for the  multi-point LHP .

\newsec{\bf Bosonization of Vertex Operator algebra}
In this way,
the multi-point LHP are expressed
in terms of the   traces
of VO.
{}From the mathematical point of view,
the vertex operators and the Corner Hamiltonian
constitute the  algebra  
\zfal ,\ \der , \uytiii .
It
acts in the direct sum of
the spaces\ ${\cal L}_{l,k}$.
Unfortunately, our "definitions" 
were rather
heuristic
than rigorous ones.
Due to  Andrews {\it et\ al.}\ \abf ,
we surely know that\ 
${\cal L}_{l,k}$\ has a
structure of
a\   ${\bf Z}$-graded space provided by the
Corner Hamiltonian.
The features 
\ \spe\ 
are characteristic
ones for a spectrum of the  grade operator 
in  
a Fock space.
Our main idea is to construct
a representation of \zfal ,\ \der ,\  \uytiii\  
in a direct sum of Fock spaces.
This representation is expected to be reducible.
However, its restriction
to an irreducible component
would provide the proper
multiplicities for 
the eigenvalues of \ ${\bf H}_C$.
A similar bosonization procedure was applied 
for calculations of
conformal blocks in the  Conformal Field Theory\
\DotsFat\ 
and lattice correlation  functions for the 
XXZ Heisenberg spin chain 
\ \japtwo ,\ \japthree .

\subsec{Bosonic representation of VO}
It is convenient for us
to perform the simple transformation of VO
\eqn\derver{
{\bf \Psi}^{\pm }(\zeta)|_{{\cal{L}}_{l,k}}=
i^{k-l} \ \sqrt{[k]}\ 
{\bf \Phi}^{\pm}(\zeta) |_{{\cal{L}}_{l,k}}
\ .}
The operators  \ ${\bf \Psi}^{\pm}$\ 
also constitute the associative 
quadratic algebra.
It can be  easily verified 
by substitution of \derver\ into \zfal .  
In particular, the commutation
relation of  \ ${\bf \Psi}^+(\zeta_1)$\
and   \ ${\bf \Psi}^+(\zeta_2)$\ has the form:
\eqn\plus{{\bf \Psi}^{+}(\zeta_1){\bf \Psi}^{+}(\zeta_2)=
R(\zeta_1\zeta^{-1}_2)\ 
{\bf \Psi}^{+}(\zeta_2){\bf \Psi}^{+}(\zeta_1)\ ,}
where the function\ $R(\zeta)$\ is given by\ \jdut  .
Our first task is to carry out the bosonization of 
\ ${\bf \Psi}^+(\zeta)$\ satisfying this simple  equation. 
To do it we 
follow the procedure developed in \refs{\singordon};
Let us introduce the operator valued function \ $\varphi(\zeta)$\
obeying the commutation relation:
\eqn\sta{
[\varphi(\zeta_1),\varphi(\zeta_2)]=
-\ln R\big(\zeta_1\zeta^{-1}_2\big)\ .}
The quasi-periodicity of
\ $R(\zeta)$\ under the substitution 
\ $\zeta\rightarrow e^{2\pi i}\zeta$\
implies the following  Laurent  decomposition  for
\ $\varphi(\zeta)$\ : 
\eqn\phdiee{
\varphi(\zeta)=-\sqrt{\frac{r-1}{2r}}\ \big({\cal Q}-i
{\cal P}\ {\rm ln}\zeta\big)
+i\sum_{\scriptstyle m\in {\bf Z}\atop
\scriptstyle  m\neq 0}\frac{\beta_m}{m}\ \zeta^{-m}\ .}
In order to provide\  \sta ,  we specify the non-vanishing
commutation 
relations for the operators \ ${\cal P, Q},\beta_n$\ as being 
\eqn\comr{\eqalign{
&[\beta_m,\beta_n]=
m\  \frac{[m]_x [(r-1)m]_x}{[2m]_x [rm]_x}\ \delta_{m+n,0}\ ,\cr
&[{\cal P,Q}]=-i\ ,
}}
where the notation
$$ [m]_x=\frac{x^m-x^{-m}}{x-x^{-1}}\ $$
is used \foot{It should not be confused with the symbol
\ $[u]$\ \oiuy .}.
With these definitions the  operator \ $\varphi(\zeta)$\ 
satisfies the commutation relations \sta. 
It is easily  checked,  since  
the function \ $R(\zeta)$\ \jdut\ 
can be rewritten in the form:
\eqn\loiu{R(\zeta)=\zeta^{\frac{r-1}{2r}}\ \exp \biggl\{
\sum_{\scriptstyle m\in {\bf Z}\atop
\scriptstyle  m\neq 0}
\ \frac{[m]_x [(r-1) m]_x}
{ m\  [2m]_x [ r m]_x}\ \zeta^{m}\ \biggr\}\ .}
The Heisenberg algebra \comr\ is 
represented in Fock spaces. In the usual fashion,
one defines
the Fock space \ ${\cal{F}}_P$\ as the  module
generated by action of the creation operators \ $\beta_{-n}, n>0$\ 
on the highest weight vector \ ${\bf v}_P$\ 
\eqn\kiuy{\beta_n{\bf v}_P=0\ ,\  n>0\ ;\ 
\ \ \ {\cal P}\ {\bf v}_P=P\ {\bf v}_P\ .}
Namely, \ ${\cal{F}}_P$\
is covered by vectors  
$\ \beta_{-n_1}
\beta_{-n_2}\dots \beta_{-n_j}{\bf v}_P\ ,\ \ n_m> 0\  .$ 
It is endowed  with  the structure of 
a ${\bf Z}$-graded module
by the  operator:
\eqn\grad{
{\bf H}_C=\sum_{m>0}\ \frac{[2m]_x  [r m]_x}{[m]_x [(r-1) m]_x}
\ \beta_{-m}\beta_{m}
+\frac{{\cal P}^2}{2} -\frac{1}{24}\ .}
Its eigenvalues
in  ${\cal F}_P$ have the
form\ \juyt\  and we would like to identify\ \grad\   
with the Corner Hamiltonian.
As it follows from\ \juyt ,
only the Fock spaces
\eqn\hytra{
{{\cal F}}_{ l,k}\equiv {\cal F}_{ \frac{rl-(r-1)k}{\sqrt{2(r-1)r }}}
\ ,}
with\   $l$\  and\  $k$\  being integers,
are relevant in our construction.
 
Now, let us introduce
the following bosonic representation:
\eqn\opem{
{\bf \Psi^{+}}(\zeta)=
e^{i\varphi(\zeta)}:\,  {{\cal F}}_{l,k}\rightarrow
{{\cal F}}_{ l,k+1}\ .}
Then the relation \ \plus\   is  satisfied
immediately.
In order to describe  
the bosonic representation for the  vertex operator
${\bf \Psi}^-(\zeta)$,
we need the
notations:
\eqn\opuy{U(\zeta)=e^{i\varphi(\zeta)}\ ,\ \ \ \ 
\bar{U}(z)=e^{-i(\varphi(z x)+\varphi(z x^{-1})) }\ .}
Define also  the operator
\ $F(z,{\cal P})$\  depending on the "zero" mode\ ${\cal P}$
\eqn\liu{F(z,{\cal P})=z^{-\omega-\frac{r-1}{r}} 
x^{r\omega(\omega+1)-\omega}\ 
\frac{E\big( x^{1-2r\omega} z,x^{2r}\big)} 
{E\big(xz^{-1},x^{2r}\big)}\ ,}
where  \ $\omega=\sqrt{\frac{2r-2}{r}}\ {\cal P}$
and \ $E(z,q)$ \ is given by \ \trassa .
We propose the following bosonization prescription:
\eqn\gtrs{
{\bf \Psi}^{-}(\zeta)=
\eta^{-1}\oint_{{\cal C }}\frac{d z}{2\pi i z} \ U(\zeta) \bar{U}(z)\ 
F\big(z\zeta^{-1},{\cal P}\big)=
\eta^{-1}\oint_{{\cal C}} \frac{d z}{2\pi i z}
\ F\big(\zeta z^{-1}, -{\cal P}\big)\  \bar{U}(z) U(\zeta)\  .}
Here the  anti-clockwise  contour \ ${\cal C}$\ 
encloses the poles\ $z=\zeta x^{1+2rm}\ (m=0,1,2,..)$\ 
of the integrand
and \ $\eta$ is some constant,
which we are going to specify later.
It is significant that  the integrand 
in\  \gtrs\ is  a single-valued 
function of \ $z$\ on the complex plane,
so the 
integration contour is closed. 
Therefore, the action of the 
operator \gtrs \
is well defined for an  arbitrary Fock space \ ${\cal{F}}_P$.
We claim that; 
\par 
{\it The operators \ \opem ,\ \gtrs
\eqn\kijnh{{\bf \Psi}^{\pm}(\zeta):
\ {\cal F}_{l,k}\to{\cal F}_{l,k\pm 1}\ ,}

satisfy the
commutation relations
\eqn\hdfasa{ {\bf \Psi}^{a}(\zeta_1) {\bf\Psi}^b(\zeta_2)
|_{{\cal F}_{l,k}}
=\sum_{c+d=a+b}{\bf U}
\left[\matrix{k+a+b&k+c\cr k+b  &k}
\biggl|\zeta_1\zeta_2^{-1}   \right]\
{\bf \Psi}^d (\zeta_2)
{\bf \Psi}^c(\zeta_1)|_{{\cal F}_{l,k}}\ .}    }
The coefficients \ ${\bf U}$ \hdfasa\  are connected
with the  Boltzmann weights\ \iu \  via  the simple transformation 
\eqn\dert{{\bf U}
\left[\matrix{m_1&m_2\cr m_4  &m_3}
\right]=i^{m_4-m_2}\ \sqrt{\frac{[m_4]}
{[m_2]}}
\ {\bf W}\left[\matrix{m_1&m_2\cr m_4  &m_3}
\right]\ .}
It is apparent that the commutation relations \ \hdfasa\ 
are equivalent to \ \zfal\ if
VO \ ${\bf \Psi}^{\pm}$\  and\  ${\bf \Phi}^{\pm}$\  are
related as in\ \derver .
The explicit form of the matrix \ \dert\ 
is presented in Appendix A.
We  give the flavour
of the  proof\ 
\hdfasa\   in  Appendix B.
It is easy to see that the operators
\opem ,\ \gtrs\ obey  the proper commutation
relation\  \der\   
with the Corner Hamiltonian \grad .
One can also check 
that they 
satisfy the  normalization condition:
\eqn\poiuy{
\biggl\{\ {\bf \Psi}^-(x^{-2}\zeta){\bf \Psi}^+(\zeta)-
{\bf \Psi}^+(x^{-2}\zeta)
{\bf \Psi}^-(\zeta)\biggr\}\Big|_{{\cal F}_{l,k}}=
A\ (-1)^{k-l}\ [k]\ { \bf I}\  \Big|_{{\cal F}_{l,k}}\   ,}
where the numerical constant \ $A$\ is equal to
\eqn\plsi{ A=\eta^{-1}\ \sqrt{1-z}|_{z\to 1}\
x^{\frac{1-r}{2r}}
\  \frac{\sqrt{(x^{2},x^{2r})_{\infty}(x^{2r-2},x^{2r})_{\infty}}}
{(x^{2r},x^{2r})_{\infty}}\ .}
The occurrence of the  multiple\ $\sqrt{1-z}|_{z\to 1}$\ 
is connected with a fact  that we
do not use the  normal ordered forms of the exponents
 \ \opuy .
The normal ordering of \ $U(\zeta)$\ 
produces  only  an additional
multiple factor
which is a  non zero and finite constant for \ $0<x<1$.
At the same time, the  normal ordering of \ ${\bar U}(z)$\ 
gives rise to
the factor
\ $\sqrt{1-z}|_{z\to 1}$.
It can be cancelled out  by a suitable choice of
\ $\eta$. 
We specify this constant in such a way that
$$A=i \ ,$$
and \ \poiuy\ 
would be equivalent
to the normalization condition
\uytiii . 

\subsec{Felder complex}
Now we should concentrate our attention 
on the description of 
\ ${\cal{L}}_{l,k}$\ in terms of the bosonic Fock spaces.
Recall, the Corner Hamiltonian was 
identified with the grade operator \grad .
Although the eigenvalues of \grad\
coincide with ones for the Corner Hamiltonian,
their multiplicities 
are different.
Hence, the spaces \ $
{\cal{F}}_{l,k}$ and \ ${\cal{L}}_{l,k}$\
cannot be  
identified directly.
More sophisticated treatment shows
that the bosonic representation
of the VO algebra
is reducible.
To construct the proper 
space of states 
one has 
to extract an irreducible component
by throwing out some
states from the Fock spaces. Explicitly, the 
procedure reads as follows; Introduce the notation
\eqn\bart{
\bar{V}(z)=e^{-i(\phi(zx)+\phi(zx^{-1}))}
}
with \ $\phi(\zeta)$\ given by
\eqn\phitru{
\phi(\zeta)=
\sqrt{\frac{r}{2(r-1)}}\
\big({\cal Q}-i{\cal P}\ {\rm ln}\zeta\big)
-i\sum_{\scriptstyle m\in {\bf Z}\atop
\scriptstyle  m\neq 0}\frac{\alpha_m}{m}\ \zeta^{-m}}
and
\eqn\dr{[(r-1)m]_x\ \alpha_m=[rm]_x\ \beta_m\ .}
A central place in the analysis belongs to 
the operator
\eqn\cre{
X={\tilde \eta}^{-1}\ \oint\frac{dz}{2\pi i z }\ 
\bar{V}(z)\ {\tilde F}(z,{\cal P})\ ,}
where
$$ {\tilde F}(z,{\cal P})
=z^{-{\tilde \omega}-\frac{r}{r-1}}\  
x^{(r-1){\tilde\omega}({\tilde\omega}+1)+{\tilde\omega}}\ 
\frac{E\big( x^{-1-2(r-1){\tilde\omega}} z,x^{2r-2}\big)} 
{E\big(x^{-1}z^{-1},x^{2r-2}\big)}\ ,$$
with\ ${\tilde \omega}=-\sqrt{\frac{2r}{r-1}}{\cal P}$\ 
and the constant \ ${\tilde \eta}$\ provides the regularization
of  \ $\bar{V}(z)$. 
The operator \ \cre\ defines
the following
maps:
\eqn\jugh{\eqalign{
X_{2j}=X^l&:\ {\cal{F}}_{l-2j(r-1),k}
\rightarrow {\cal{F}}_{-l-2j(r-1),k}\ ,\cr 
X_{2j+1}=X^{r-1-l}&:
\ {\cal{F}}_{-l-2j(r-1),k}
\rightarrow {\cal{F}}_{l-2(j+1)(r-1),k}\  }}
for\ $j\in{\bf Z},\ 1\leq l\leq r-2,\ 1\leq k\leq r-1$.
As a result, we can construct 
the infinite chain:
\eqn\fel{\matrix{&&    X_{-2}
         &                   & X_{-1}       &                & 
X_{0}           &                & X_{1}       &  & \cr
           &    ...&
\longrightarrow 
&{\cal{F}}_{2r-2-l,k}&\longrightarrow&{\cal{F}}_{l,k}&
\longrightarrow&{\cal{F}}_{-l,k}&\longrightarrow&...&}\ .}
We claim
that it is the Felder resolution 
\refs{\felder}.
In other words,

{\it I.\   The chain of maps\    \fel\ is a BRST complex
for any 
\ $0<x<1$ , i.e.}
\eqn\brst{
X_{j }X_{j-1}
=0\ ;}

{\it II.\  The cohomologies
of the complex turn out to be non-trivial only
for\ $j=0$}
\eqn\coh{
Ker\ X_{j}\Big/Im\ X_{j-1}=0\ ,  \ {\rm if} \ j\neq 0 \ .}
The first statement can be checked by
the direct calculations\ \las .
The arguments in favour 
of\  \coh \
are based on the fact 
that the operator \ \cre\ 
commute with\ ${\bf H}_C$\ \grad .  
So, the cohomology spaces
are 
\ $\bf Z$-graded, as well as\ ${\cal F}_P$.
Let us consider the
finite-dimensional eigensubspaces of the  operator \ ${\bf H}_C$\
in \ $Ker\ X_{j}\Big/Im\ X_{j-1}$.
The restriction of\  \fel\ \ on these subspaces
provides BRST complexes involving only
a finite number of the maps.
Dimensions of the cohomology spaces
for such complexes  
are
integers and would
not depend
on the continuous parameter \ $ x $.
Therefore, it is enough to find them for
an arbitrary 
\ $0<x<1$. 
The statement\ \coh\ 
surely holds 
at the limit \ $x\rightarrow 1$,
where our construction 
turns out to be the
Feigin-Fuks-Dotsenko-Fateev bosonization
\ \fei , \DotsFat .
In addition, we conclude that 
the spectrum of \ ${\bf H}_C$\  
in the space
\eqn\gspace{
{\cal L}_{l,k}=Ker\ X_{0}\Big/Im\ X_{-1}=
Ker_{{\cal{F}}_{l,k}}\
X^{l}\ \Big/Im_{{\cal{F}}_{2r-2-l,k}}\ X^{r-1-l}}
coincides with the spectrum of the
grade operator $L_0-\frac{c}{24}$\ in the
corresponding 
irreducible
representation of the  Virasoro algebra. 

To complete the identification of the
factor-space \gspace \ with \ ${\cal{L}}_{l,k}$\ 
one notes that VO satisfy the commutation relations:
\eqn\invsa{
{\bf \Psi}^\pm(\zeta)\  X_j
= X_j\ {\bf \Psi}^{\pm}(\zeta)\ .}
So they are BRST invariant operators.

The BRST properties and the structure
of the
complex \fel\ lead to the following computation procedure
of traces over  
\ ${\cal L}_{l,k}$\ in terms of  traces over the Fock
spaces \ \felder . Introduce the notation:
\eqn\lokj{
{\cal{O}}_n= 
{\bf \Psi}^{-\sigma_1}(x^{-2}\zeta_1)...
{\bf \Psi}^{-\sigma_{n-1}}(x^{-2}\zeta_{n-1})\ 
{\bf \Psi}^{\sigma_{n-1}}(\zeta_{n-1})...
{\bf \Psi}^{\sigma_1}(\zeta_1)\  .}
Then, due to the commutation relations \invsa\
the following diagram 
is a  commutative one:
\eqn\diagr{\eqalign{
\matrix{
&{\  }&X_{-2}& {\ }&X_{-1}&{\ }&X_{0}&{\ }&X_{1}&{\ }\cr
&...&\longrightarrow&\ \ \ \ {\cal F}_{-l+2r-2, k}
&\longrightarrow &{\cal F}_{l,k} &{\longrightarrow }
&{{\cal F}_{-l,k} }&{\longrightarrow }& ...\cr
&{\ }&{\ }&\ \ \ \ \ \ \biggl\downarrow
x^{4{\bf H}_C}{\cal{O}}_n&{\ }
&\ \ \ \ \ \ \ \biggl\downarrow x^{4{\bf H}_C}{\cal{O}}_n&{\ }&
\ \ \ \ \ \ \biggl\downarrow x^{4{\bf H}_C}{\cal{O}}_n&
{\ }&{\ }\cr
&...& \longrightarrow  &\ \ \ \ {\cal F}_{-l+2r-2, k}             
&\longrightarrow &{\cal F}_{l,k} &{\longrightarrow }
&{{\cal F}_{-l,k} }&{\longrightarrow }& ...\cr
&{\ }&X_{-2}& {\ }&X_{-1}&{\ }&X_{0}&{\ }&X_{1}&{\ }\cr
}\ \ \ \ .}}
The complex is exact excluding the \ $j=0$\  term,
hence the trace of the
operator \ $\ x^{4 {\bf H_C}}{\cal{O}}_n$\
over the space\ \gspace\ 
equals to the following
alternating sum:
\eqn\trass{\eqalign{
&Tr_{{\cal L}_{l,k}}\Big[\  x^{4 {\bf H_C}}{\cal{O}}_n\ \Big] 
=Tr_{Ker_{{\cal{F}}_{l,k}}\
X^{l}\big/Im_{{\cal{F}}_{2r-2-l,k}}\ X^{r-1-l}}
\Big[\  
x^{4 {\bf H_C}}{\cal{O}}_n\ \Big]= \cr
&\sum_{j=-\infty}^{\infty}Tr_{{\cal{F}}_{l-2j(r-1),k}}
\Big[\  x^{4 {\bf H_C}}{\cal{O}}_n\ \Big] \
-\sum_{j=-\infty}^{\infty}Tr_{{\cal{F}}_{-l-2j(r-1),k}}
\Big[\  x^{4 {\bf H_C}}{\cal{O}}_n\ \Big] \ .}}
In the simplest case\ ${\cal{O}}_1=1$\ 
the formula \trass\ gives
\eqn\chari{\eqalign{
Tr_{{\cal L}_{l,k}}&\Big[\  x^{4 {\bf H_C}}\ \Big]
=\sum_{j=-\infty}^{\infty}Tr_{{\cal{F}}_{l-2j(r-1),k}}
\Big[\  x^{4 {\bf H_C}}\ \Big] \
-\sum_{j=-\infty}^{\infty}
Tr_{{\cal{F}}_{-l-2j(r-1),k}}\Big[\  x^{4 {\bf H_C}}\ \Big]= \cr
&x^{\frac{1}{6}}\ (x^4;x^4)^{-1}_\infty\ 
\biggl\{\ \sum_{j=-\infty}^{\infty}  
x^{ \frac{(rl-(r-1)k+2(r-1)rj)^2}{(r-1)r}} \
-\sum_{j=-\infty}^{\infty}
x^{\frac{(rl+(r-1)k+2(r-1)r j)^2}{(r-1)r}}\ \biggr\} \ .}}
It coincides with the character \ $\chi_{l,k}(x^4)$\ 
\  \char  ,
due to the Jacobi identity: 
$$E(z,q)=\sum_{m=-\infty}^{+\infty}\ (-1)^m
\ q^{\frac{(m-1)m}{2}}\ z^m\ .$$

\newsec{Integral representation for  multi-point LHP}
\subsec{Calculation of traces}
In this section we work out an integral representation
for  LHP by using the results of 
the bosonization procedure. In terms of VO 
\ ${\bf \Psi}^{\pm}(\zeta)$\ the equation\ 
\hydtr \ can be rewritten as
\eqn\hydtor{\eqalign{&P_{k_1,...k_n}(\zeta_1,...\zeta_{n-1}| l)=
i^{2l-k_1-k_n}\
{\bar Z}^{-1}_l\  \prod^{n-1}_{s=2}  [k_s]^{-1}(-1)^{l-k_s}
\times\cr 
&Tr_{{\cal L}_{l,k_1}}\Big[\ x^{ 4 {\bf H}_C}\ 
{\bf\Psi}^{-\sigma_1}(x^{-2}\zeta_1)
...{\bf \Psi}^{-\sigma_{n-1}}(x^{-2}\zeta_{n-1})
{\bf \Psi}^{\sigma_{n-1}}(\zeta_{n-1})...
{\bf \Psi}^{\sigma_1}(\zeta_1)\ \Big]\ .}}
It is obvious that in the calculation of\  \hydtor \ 
we can perform the traces over the zero and 
oscillator modes separately.
The zero mode  contribution is governed by 
the structure of the complex\ \diagr\  and it was obtained 
in\ \felder . 
The trace over the oscillator part is carried out
by using the Clavelli-Shapiro technique \ \cls , \japthree  .
More explicitly, the prescription
implies  the introducing together with
\ $\beta_n$\ the oscillators
$$
\beta^*_m:\ 
\ [\beta_m,\beta^*_n]=0 \ ,
$$
satisfying  the same commutation relations as\  \comr .
Define the following operators
acting in the tensor product
\ ${\cal F}[\beta]\otimes {\cal F}[\beta^*]$\ 
of  the Fock spaces:
\eqn\comrr{\eqalign{
&b_m=
(1-x^{4m})^{-1}\  \beta_m\otimes 1+1\otimes \beta^*_{-m}\ ,
\ m>0\ ;\cr
&b_m=\beta_m\otimes 1+
({x^{4m}-1})^{-1}\ 1\otimes\beta^*_{-m}\ , \ m<0\ .}}
Notice, that we  omit the index \ $P$\ since
the result does not depend 
on the zero mode eigenvalue in this case.
For any operator  
\ ${\cal{O}}[\beta]$ \ acting in
\ ${\cal F}[\beta]$ \ one can construct
the operator
$${\cal O}[b]: 
\ {\cal F}[\beta]\otimes {\cal F}[\beta^*]
\rightarrow 
\ {\cal F}[\beta]\otimes {\cal F}[\beta^*]
$$ 
by the  replacement 
of the modes \ $\beta_m$\  with \ $b_m$\  \comrr .  
Then the "oscillator" trace over the Fock space
is expressed in terms of the vacuum expectation value 
\ $<<{\cal O}[b]>>\equiv <0|{\cal{O}}[b]|0>$\
of the  operator
\ ${\cal O}[b]$\
with respect to the  vacuum\ $|0>={\bf v}\otimes{\bf v}.$
Namely,
\eqn\tra{
Tr_{osc}\Big[\ x^{4{\bf H}_C}{\cal{O}}[\beta]\  \Big]=
(x^4;x^4)^{-1}_{\infty}\ <<{\cal{O}}[b]>>\ .}
Due to the Wick theorem, the 
expectation value   
of a product of the exponential
operators 
\ \opuy\    is factorized
into
the  two point functions
\foot
{These functions were calculated in the work \refs{\singordon}
(see also\  \japnews ), and
we preserved here the same notations. The prime
should not be confused with the symbol of the   derivative.}:
\eqn\func{\eqalign{
&<<U(\zeta_2)U(\zeta_1)>>= {{\cal R}}^2\ G'(\zeta_1\zeta^{-1}_2) \cr
&<<\bar{U}(\zeta_2)U(\zeta_1)>>=\eta\  {\cal R}\ {\bar {\cal R}}\ W'
(\zeta_1\zeta^{-1}_2)\cr
&<<\bar{U}(\zeta_2)\bar{U}(\zeta_1)>>=
\eta^2\   {\bar {\cal R}}^2\ \bar{G}'
(\zeta_1\zeta^{-1}_2)\ .}}
The calculations of the functions and
constants in\ \func\ are 
straightforward, and their explicit forms  are written down
in Appendix C.
\subsec{Integral representation}
We begin with the following trace:
\eqn\juy{{\hat P}_{k,k+ 1}(\zeta_1\zeta_2^{-1}|l)=i\ 
(-1)^{l-k-1}\ {\bar Z}^{-1}_l
\ Tr_{{\cal L}_{l,k}}\Big[\ x^{4{\bf H}_C}\ 
 {\bf \Psi}^{-}(\zeta_2)
{\bf \Psi}^{+}(\zeta_1)\ \Big]\ .}
According to our analysis,
\juy\   admits the bosonic representation\  \trass\
with 
\eqn\twop{
{\cal{O}}_2=
\eta^{-1}\ \oint_{|z|=|\zeta_2|  }\frac{d z}{2\pi i z} \
F(\zeta_2 z^{-1},-{\cal P})\ {\bar U}(z)U(\zeta_2)U(\zeta_1)\ .}
First, let us derive the zero modes contribution  
into the integrand.
For the \ $j=0$\ term in the sum \trass\ one 
obtains:
\eqn\nulli{\eqalign{
x^{2{\cal{P}}^2-\frac{1}{6}}\ 
F(\zeta_2 z^{-1},-{\cal{P}})&\ 
e^{2 i \sqrt{ \frac{r-1}{r}}  (Q-i{\cal{P}} \ln{z}) }\ 
e^{-i \sqrt{ \frac{r-1}{r}} (Q-i{\cal{P}} \ln{\zeta_2}) }\ 
e^{- i \sqrt{ \frac{r-1}{r}} (Q-i{\cal{P}} \ln{\zeta_1})\  
}\biggl|_{{\cal F}_{l,k}}
=\cr
&=x^{\frac{r\omega^2}{r-1}-\frac{1}{6}}\ 
\big({\zeta_1}{\zeta_2}^{-1}\big)^{\frac{r-1}{4r}-
\frac{\omega}{2}}\ 
\big(z\zeta^{-1}_2\big)^{\omega-\frac{r-1}{r}}\ 
F(\zeta_2 z^{-1},-P)\ ,}}
where\ $\omega=\sqrt{\frac{2r-2}{r}}\ P=l-\frac{r-1}{r} k$.
All other terms
can be found 
by the substitution
\ $l\rightarrow \pm l-2j(r-1)\ , \ j\in {\bf Z}$\
into\ \nulli  . Then, performing the sum over \ $j$\
we get the zero 
mode contribution:
$$(x^4;x^4)_{\infty}\
\big(\zeta_1\zeta_2^{-1}\big)^{\frac{r-1}{4r}-\frac{\omega}{2}}\ 
O_{l,k}\big(\zeta_1\zeta_2 z^{-2}\big)\ 
(z\zeta_2^{-1})^{\omega-\frac{r-1}{r}}\ 
F(\zeta_2 z^{-1},-P)\ ,$$
where 
\eqn\kio{\eqalign{
O_{l,k}(\zeta)=
(x^4;x^4)^{-1}_{\infty}&\ x^{-\frac{1}{6}}\ 
\Big\{ x^{\frac{(rl-(r-1)k)^2}{(r-1)r}}\ 
E(-\zeta^{1-r} 
x^{4(rl-(r-1)k+(r-1)r)}, x^{8 (r-1)r})-\cr
&\zeta^l\ x^{\frac{(rl+(r-1)k)^2}{(r-1)r}}\ 
E(-\zeta^{1-r} 
x^{-4(rl+(r-1)k-(r-1)r)}, x^{8 (r-1)r})\Big\}\ .}}
The trace over the oscillator modes
is provided
by  the Wick theorem and
\  \func .
As a result, we derive the integral 
representation  for \juy :
\eqn\juyaaa{\eqalign{ &{\hat P}_{k,k+1}(\zeta_1\zeta_2^{-1};l)=
{\bar Z}^{-1}_l\ i  
\ (-1)^{l-k-1}\ {{\cal R}}^2\ {\bar {\cal R}}\ G'(\zeta_1\zeta^{-1}_2)
\ \big(\zeta_1\zeta_2^{-1}\big)^{\frac{r-1}{4r}-\frac{\omega}{2}}
\times\cr
&\oint_{|z|=|\zeta_2|  }\frac{d z}{2\pi i z} \
O_{l,k}(\zeta_1\zeta_2 z^{-2})\ 
W'(\zeta_1z^{-1})W'(\zeta_2 z^{-1})
\ (z\zeta_2^{-1})^{\omega-\frac{r-1}{r}}\ F(\zeta_2 z^{-1},-P)\ . }}
The explicit form of the  functions and constants appearing here
is rather complicated (see Appendix C).
The function\  \juyaaa\ 
represents
LHP on the lattice with more general
dislocation, than
one considered before:
\vskip 0.5cm
\centerline{
\epsfxsize 4.0 truecm
\epsfbox{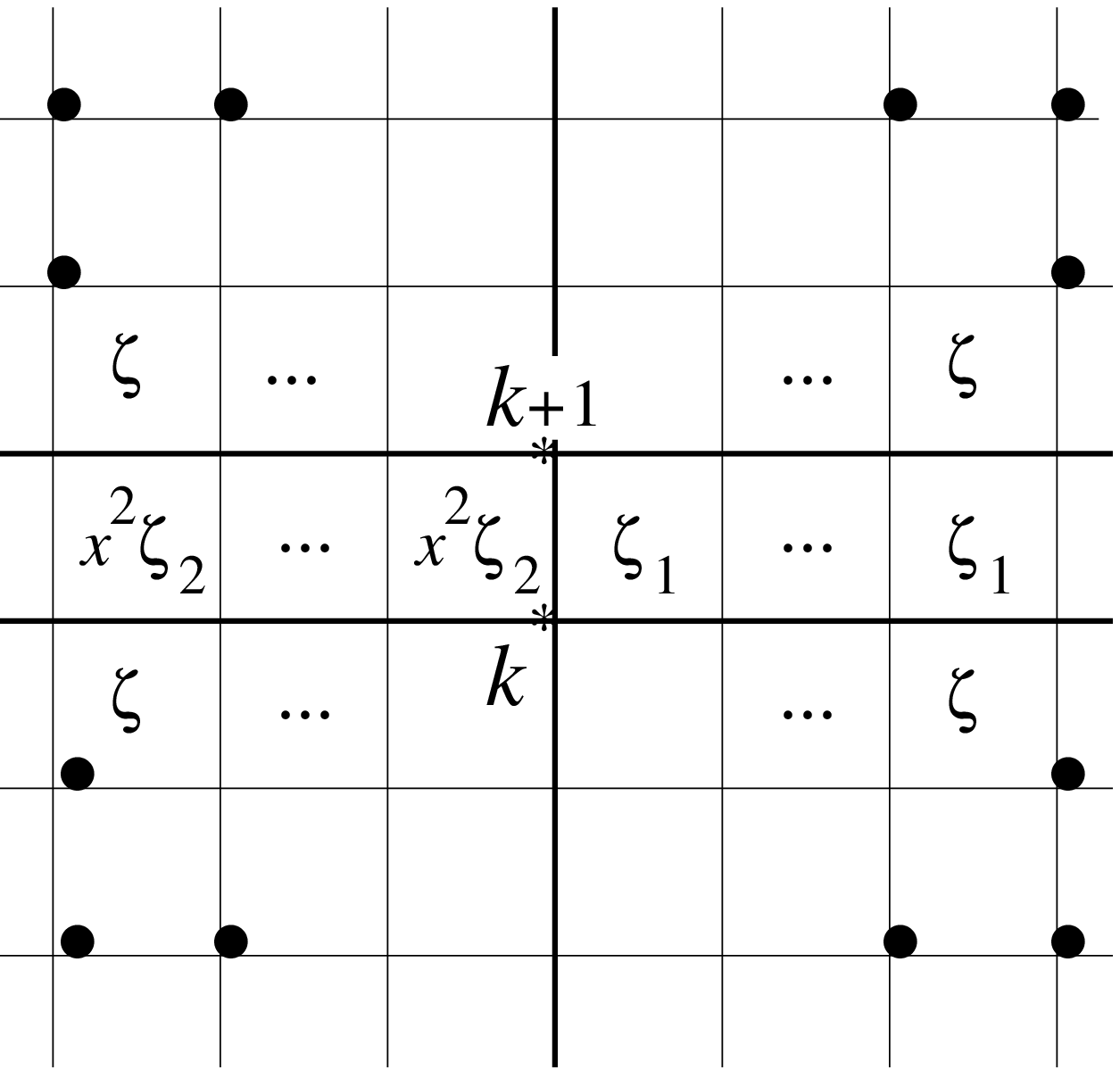}}
\vskip 0.5cm
\par\noindent
In the case \ $\zeta_1=x^2\zeta_2$\ 
the expression \ \juyaaa\ is simplified drastically.
Indeed, using the obvious  relation
$$
W'(\zeta x)W'(\zeta x^{-1})=\bar{G}'^{-1}(\zeta)\ ,
$$ 
it is possible to show
that the two-point LHP\ \uytr\  are  given by:
\eqn\oiuys{\eqalign{P_{k,k\pm 1}(l)=(-1)^{l-k}&\ {\bar Z}^{-1}_l\ 
\frac{ (x^2,x^2)^3_{\infty}}{E(x^{2r-2},x^{2r})}\
x^{r\omega(\omega-1)+\frac{r-1}{r}}\times\cr 
&\oint_{|z|=x^{\mp 2}}\frac{d z}{2\pi i z} \  O_{l,k}( z^{-2})\ 
\frac{E(z^{-1} x^{2rl-2(r-1)k}, x^{2r})}{E(z x^2,x^2)}\ .}} 
In Appendix D, we prove that
these functions coincide with\ \mssy .

In a  similar manner, 
we have  obtained the integral 
representation for the
multi-point Local
Height Probabilities:
\eqn\bssfd{\eqalign{&
P_{k_1,...k_n}(\zeta_1,...\zeta_{n-1}
|l)=(-1)^{\frac{k_n-k_1+n-1}{2}}\ [k_1]\ {\bar Z}^{-1}_{l}
\times\cr& 
\prod_{s=1}^{n-1}B^{-1}_{l,k_s}
\prod_{s<j}B^{-1}(\zeta_s \zeta^{-1}_j)
\prod_{s=1}^{n-1}\Big\{
\oint_{{\cal C}_s} \frac{d z_s}{2\pi i z_s}\Big\}\
O_{l,k_1}\big(\prod_{s=1}^{n-1}\zeta_s^2 z_s^{-2} \big)
\times\cr&
\prod_{s<j}B^{-1}(z_s z^{-1}_j x^{2k_{s+1}-2k_s})
\prod_{j<s}B(\zeta_j  z^{-1}_s )
\prod_{j>s}B(\zeta_j z^{-1}_s  x^{2k_s-2k_{s+1}})
\prod_{s=1}^{n-1}B_{l,k_s}(z_s\zeta^{-1}_s)\ ,}}
where 
$$B(z)=
\frac{E(z,x^{2r})}{E(z,x^2)}\ ,$$
$$B_{l,k}(z)
=\frac{E(z x^{-2rl+2(r-1) k} ,x^{2r})}
{E(z,x^2)}\ ,$$
$$B_{l,k}=\frac{E(x^{-2rl+2(r-1) k},x^{2r})}{(x^2;x^2)^3}\ ,$$  
and  the functions \
$[k],\ {\bar{Z}}_l,\ E(z,q),\  O_{l,k}(\zeta)$
\ are given
by\ \oiuy ,\ \zl ,\ \trassa ,\ \kio , respectively.
The integration contour\ ${\cal C}_s,\ (s=1,...n-1)$\ in\ \bssfd\
goes anti-clockwise and encloses the
following poles:
$$\eqalign{z_s=\biggl\{\matrix{
&\zeta_j x^{2m+1-k_{s+1}+k_s}\ ,\ \ \ \ \ \ \ 
\ \ \ \ \ \ \  \ \ \ \ \ \ \    & \cr
&z_j x^{r(2m+1)+(r-2)(k_{j}-k_{j+1})}\ \  (j<s)\ , & 
\   z_j x^{r(2m+1)+(r-2)(k_{s+1}-k_s)}\ \  (j>s)} \ ,}$$
where\ $m=0,1,2,...$ .
\newsec{Conclusion}
In summary, 
we would like 
to make some comments on an algebraic 
interpretation of the presented results. 
Our construction
leads to a natural conjecture
that the space \ ${\cal{L}}_{l,k}$\
can be treated as the irreducible representation
of some infinite-dimensional  
algebra. Since at the limit \ $x\rightarrow 1$\ 
the  bosonization
turns out to be
the Feigin-Fuks-Dotsenko-Fateev procedure\ \fei ,  \DotsFat ,
then
we have called the algebra
as  the deformed Virasoro  one  \ $Vir_{c,x}$\
in the work \ \lp .
The explicit basis for  \ $Vir_{c,x}$\
was found in the works \refs{\yap},\  \ff
\foot{ J. Shiraishi  {\it et al.}\ \yap\ use the notations
$ p=x^{-2}, \ q=x^{-2r},\ \ t=q p^{-1}=x^{2-2r} \ . $}.
It can be described in terms of
the  oscillators \ $\lambda_m$ \ : 
\eqn\ewq{[(r-1)m]_x\ \alpha_m=[rm]_x\ \beta_m=
(x-x^{-1})^{-1}\ m\ \lambda_m\ .}
If one introduces the field
\eqn\hy{\Lambda(z)=x^{\sqrt{2r (r-1)}{\cal P}}
\  :{\rm \exp}
\Biggl(-\sum_{m\not= 0} \lambda_m z^{-m}\Biggr):\ ,}
then the generating function for the
elements of  \ $Vir_{c,x}$\
has the form \ \fr , \  \yap ,\ \ff :
\eqn\jd{T(z)=\Lambda(z x^{-1})+\Lambda^{-1}(z x)\ .}
This field commutes
with the operators \ \jugh\ 
and obeys
the relation
\eqn\bas{\eqalign
{f(\zeta z^{-1})\ 
T(z)& T(\zeta)-f(z\zeta^{-1})\  T(\zeta) T(z)=\cr
& 2\pi\ (x-x^{-1})\  [r-1]_x\ [r]_x\
\big(\ \delta(\zeta   z^{-1} x^{-2})-
\delta(\zeta  z^{-1}  x^2)\ \big)\  ,}}
where  we denote 
\ $\delta(z)=\frac{1}{2 \pi }\sum_{m=-\infty}^{+\infty} z^m\ $\
and
$$f(z)=
(1-z)^{-1}\  \frac{(z x^{2r};x^4)_{\infty}\
(z x^{2-2r};x^4)_{\infty}}
{(z x^{2r+2};x^4)_{\infty}\ (z x^{4-2r};x^4)_{\infty}}\ . $$
The limiting \ $x\to1$\ behavior of\ \jd\ 
is governed by the expansion:
\eqn\dg{T(z)=
2+r(r-1)\ (x-x^{-1})^2\  \Biggl(   \sum_{m=-\infty}^{+\infty}
L_m z^{-m}-
\frac{c}{24 }\Biggr)+O((x-x^{-1})^4)\ ,}
and \ \bas\ 
gives us the  Virasoro algebra
commutation relations  for  \ $L_m$\
with 
the central charge\ $c$\ \hytr\ .

Remind, that the operators \ ${\bf \Psi}^{\pm}$\ 
\opem \ , \gtrs\ intertwine the irreducible
representations \ ${\cal{L}}_{l,k}$\
with the same first index \ $l$.
In the terminology of the work \ \japtwo , they are the  first
kind VO.
It would
be rather natural to introduce  the second kind VO,
which intertwine the spaces \ ${\cal{L}}_{l,k}$\ 
and \ ${\cal{L}}_{l\pm 1,k}$\ . 
Their bosonization
is  easily 
performed in terms of the  oscillators
\ $\alpha_m $\ \dr .
Then the "Corner" 
space of states
\foot{
The lattice analogue of
the Angular Quantization  space\ \singordon .}\
\eqn\kiuy{\pi_Z\equiv
\sum_{1\leq l< k\leq r-1}\ \oplus {\cal L}_{l,k}\ }
of the ABF model
at the regime III
can be   treated as the irreducible representation
of the complete algebra of both kinds VO.
Notice,  that we restrict the sum
\ \kiuy\ by \ $1\leq l< k\leq r-1 $, since
$${\cal L}_{l,k}\simeq{\cal L}_{r-1-l,r-k}\ .$$

The integrability of the  ABF
model
implies an existence of
an infinite dimensional Abelian symmetry. 
The
deformed Virasoro algebra provides
another type
of the  symmetry  
which is usually referred as the dynamical one\ \refs{\japtwo}
(see also \ \note).
The generators of such  symmetry 
do not mutually  commute,
but create the  space of states of the  model.
Other well known examples of the   dynamical symmetries 
were produced by the Conformal 
Field Theory\ \CFT.
{}From the  algebraic point of view 
the symmetry corresponding to the
deformed Virasoro algebra         
is  very similar to the conformal invariance,
although the simple  geometrical meaning of 
the latter
seems to be lost.

\hskip1.5cm

\centerline{\bf Acknowledgments}
\hskip0.5cm

We
would like to thank
M. Jimbo, A. Leclair, T. Miwa and A.B. Zamolodchikov for
interesting discussions.
We
are
grateful to
Research Institute for
Mathematical Sciences
and Prof. T. Miwa
for kind invitation and hospitality.
Ya.P. also acknowledge
A.A. Belavin, B.L. Feigin and  M. Lashkevich
for helpful comments.
This work is supported in part by NSF  (S.L.)
and RFFI (95-02-05985) (Ya.P.)   grants.
\newsec{Appendix A}
For convenience, we collect in this Appendix two
different forms of the matrix\ ${\bf U}$\ \dert .
In terms of the  parameters
\ $u$\ and\ $p$,
it reads:
\eqn\iussa{\eqalign{
&{\bf U}
\left[\matrix{m\pm 2&m\pm 1\cr m\pm1  &m}
\right]=R\ ,\cr
&{\bf U}
\left[\matrix{m&m\pm 1\cr m\pm1  &m}
\right]=R\ \frac{[m\pm u]}{[1-u]
[m]}\ ,\cr
&{\bf U}
\left[\matrix{m&m\pm 1\cr m\mp1  &m}
\right]=-R\
\frac{[u][m\mp 1]}{[1-u][m]}
\ .}}
Via the variables \ $\zeta$\ 
and\ $x$\  \jdufyt , 
the same functions
are
given  by the formulas:
\eqn\iuy{\eqalign{
&{\bf U}
\left[\matrix{m\pm 2&m\pm 1\cr m\pm1  &m}
\right]=R\ ,\cr
&{\bf U}
\left[\matrix{m&m\pm 1\cr m\pm1  &m}
\right]=R\  \zeta^{\frac{(r-1)(\mp m-1)}{r}}
\ \frac{E(x^{2r-2},x^{2r})\ E(x^{\mp 2(r-1) m} \zeta,x^{2r})}
{E(x^{2r-2} \zeta,x^{2r})\ E(x^{\mp 2(r-1) m} ,x^{2r})}\ , \cr
&{\bf U}
\left[\matrix{m\pm 2&m\mp 1\cr m\pm1  &m}
\right]=-R \ x^{\mp \frac{2(r-1) m}{r}}\ \zeta^{\frac{1-r}{r}}\ 
\frac{E(x^{2(r-1)(\pm m+1)},x^{2r})\ E( \zeta,x^{2r})}
{E(x^{2r-2} \zeta,x^{2r})\ E(x^{\mp2(r-1)  m} ,x^{2r})}
\ .}}
The function\ $R=R(\zeta)$\ is specified by the
equation\ \jdut . 

\newsec{Appendix B }
We
show
that bosonic operators\ \opem , \gtrs\   satisfy\  \hdfasa\ 
by working out  the example of the commutation relation
\eqn\mnbvo{\eqalign{
{\bf \Psi}^{-}(\zeta_1)
{\bf \Psi}^{+}(\zeta_2)&|_{{\cal{L}}_{l,k}}=
{\bf U}
\left[\matrix{k&k-1\cr k+1&k}\biggl|\  \zeta_1\zeta^{-1}_2
\ \right]
{\bf \Psi}^{+}(\zeta_2){\bf \Psi}^{-}(\zeta_1)|_
{{\cal{L}}_{l,k}}+\cr &
{\bf U}
\left[\matrix{k&k+1\cr k+1&k}\biggl|\  \zeta_1\zeta^{-1}_2
\ \right]\
{\bf \Psi}^{-}(\zeta_2){\bf \Psi}^{+}(\zeta_1)|_
{{\cal{L}}_{l,k}}\ .}}
To simplify the formulas, let us return 
to the parameterization\ \jdufyt .
In terms of the variables\ $p$\ and $u$\ the
function \ $F(\zeta,{\cal P})=F[u,\omega]$ \ \liu\ reads:
\eqn\ijhg{F[u,\omega]=\frac{[1/2+u-r\omega]}{[1/2-u]}\ ,}
where \ $\omega=\sqrt{\frac{2r-2}{r}}\ {\cal P}$ and
the symbol\ $[u]$\ is defined by\ \oiuy .
Then, the following equation
\eqn\kiuyt{\eqalign{
&F[v-u_1, r\omega-r+1]=\frac{[r-1][r\omega+u_1-u_2]
[1/2+v-u_1][1/2+u_2-v]}{[r-1+u_1-u_2][r\omega][1/2-v+u_1]
[1/2-u_2+v]}\times\cr &
\ \ \ \ \ \ F[v-u_2, r\omega-r+1]-
\frac{[r-1-r\omega][u_1-u_2][1/2+u_2-v]}{[\omega][r-1+u_1-u_2]
[1/2-u_2+v]}\ F[v-u_1,\omega]\ }}
is a simple consequence of the famous Riemann identity:
\eqn\juhg{\eqalign{[2x][2y]&[2z][2t]=
[x+y+z+t][x-y-z+t][x+y-z-t][x-y+z-t]+\cr
&[-x+y+z+t][x-y+z+t][x+y-z+t][x+y+z-t]\ .}}
By using\ 
\kiuyt , it is easy  to obtain the relation between
the exponential operators\ \opuy :
\eqn\jduhbvc{\eqalign{
&U(\zeta_1){\bar U}(z) F(z\zeta^{-1}_1, {\cal P})
U(\zeta_2)|_
{{\cal{L}}_{l,k}}=
{\bf U}
\left[\matrix{k&k+1\cr k+1&k}\biggl|\  \zeta_1\zeta^{-1}_2
\ \right]\ U(\zeta_2) U(\zeta_1)
{\bar U}(z)\times\cr&
F(z\zeta^{-1}_2, {\cal P})|_
{{\cal{L}}_{l,k}}
+{\bf U}
\left[\matrix{k&k-1\cr k+1&k}\biggl|\  \zeta_1\zeta^{-1}_2
\ \right]\
\ U(\zeta_2)
{\bar U}(z)
F(z\zeta^{-1}_1, {\cal P})
U(\zeta_1)|_
{{\cal{L}}_{l,k}}\ ,}}
here\ $\zeta_{1,2}=x^{2u_{1,2}}\ , z=x^{2v}$.
Now, to derive\ \mnbvo ,
we should integrate 
both 
parts of this equation over the variable\ $ z$.

\newsec{Appendix C. Reference formulae for trace calculations}
Here we
collect the formulae for the functions
and constants in
\ \func :
\eqn\gprim{\eqalign{
G'(\zeta)=&
\frac{(x^{2r+2}\zeta;x^{2r},x^4,x^4)_\infty\  
(x^{2}\zeta;x^{2r},x^4,x^4)_\infty}
{(x^{2r}\zeta;x^{2r},x^4,x^4)_\infty\  
(x^{4}\zeta;x^{2r},x^4,x^4)_\infty}\times\cr
&\frac{(x^{2r+6}\zeta^{-1};x^{2r},x^4,x^4)_\infty\  
(x^{6}\zeta^{-1};x^{2r},x^4,x^4)_\infty}
{(x^{2r+4}\zeta^{-1};x^{2r},x^4,x^4)_\infty\  
(x^{8}\zeta^{-1};x^{2r},x^4,x^4)_\infty}\ ,}}
\eqn\wprim{
W'(\zeta)=
\frac{(x^{2r-1}\zeta;x^{2r},x^4)_\infty \ 
(x^{2r+3}\zeta^{-1};x^{2r},x^4)_\infty }
{(x\zeta;x^{2r},x^4)_\infty \ 
(x^{5}\zeta^{-1};x^{2r},x^4)_\infty }\ ,}
\eqn\barg{
{\bar G}'(\zeta)=
\frac{(\zeta;x^2)_\infty \ 
(x^{4}\zeta^{-1};x^2)_\infty }
{(x^{2r-2}\zeta;x^{2r})_\infty \ 
(x^{2r+2}\zeta^{-1};x^{2r})_\infty }\ ,}
\eqn\mjuyl{{\bar {\cal R}}=i\ x^{\frac{r-1}{2r}}\  
\frac{(x^2;x^2)_\infty \ 
(x^{2r};x^{2r})_\infty }
{(x^{2r-2};x^{2r})_\infty \ 
(x^{2};x^{2r})_\infty }\ , }
\eqn\mjuy{{\cal R}^2=G'(1) \ .}

The functions satisfy the relations
\eqn\kiujh{\eqalign{&G'(\zeta x)G'(\zeta x^{-1})=
{W'}^{-1}(\zeta)\ ,\cr
&W'(\zeta x)W'(\zeta x^{-1})={\bar G}^{'-1}(\zeta)\ ,\cr
&G'(\zeta)=G'(x^4\zeta^{-1})\ ,\cr
&W'(\zeta)=W'(x^4\zeta^{-1})\ ,\cr
&{\bar G}'(\zeta)={\bar G}'(x^4\zeta^{-1})\ .}}

\newsec{Appendix D.}
In this Appendix 
we demonstrate that the
expressions\ \mssy ,
\ \oiuys\    for the  two-point LHP are equivalent.
The integral \ \oiuys\
is given by 
a sum of the  
residues of the  poles
situated inside the
circle\ $|z|=x^{-2}$.
The direct calculation leads to the formula:
\eqn\oy{\eqalign{&P_{k,k+1}(l)=
(-1)^{l-k}\ {\bar Z}_l^{-1}\
\frac{ x^{4 \Delta_{l,k}-\frac{c}{6} }}{E(x^{2r-2}, x^{2r})
(x^4;x^4)_{\infty}}\times\cr&
x^{r\omega(\omega-1)+\frac{r-1}{r}}
\ \Big\{J\big(x^{2rl-2(r-1)k}\big)-x^{4 r l k +2 rl}\ 
J\big(x^{-2rl-2 (r-1)k}\big)\Big\}\ .}}
Here\  $\Delta_{l,k}, \ c$\ are  defined by \ \loiuy\  and
\eqn\gty{J(m)=-\frac{m}{2} \sum_{s=0}^{2r-3}\ (-1)^s \ x^{s(s-1)} 
E(-m^2 x^{4(r-1)s+4(r-1)r)}, x^{8(r-1)r}) E'(m x^{-2s}, x^{2r})\ .}
The prime in \ \gty\ 
denotes a   derivative with respect to the first argument.
Using  simple  properties of
the elliptic function\ $E(z,q)$,
one can show that\  
$J(m)$\ satisfies the
identity:
\eqn\ui{J(x^{2r-2} m)=-m^{-1} E(-m^2 x^{4(r-1)r}, x^{8(r-1)r})
 E(m, x^{2r})+
m^{-1} J(m)\ .}
It provides the 
recursion relation
\eqn\oip{P_{k,k+1}(l)+P_{k,k-1}(l)={\bar Z}_l^{-1}\ [k]
\ \chi_{l,k}(x^4)\ ,}
where\ $\chi_{l,k}(x^4)$\ is  given  by \ \char .
It only remains to check
that the function\ $ P_{1,0}(l)$ \ obeys  the
condition\ \mjhg .
In terms of   integrals this  means:
\eqn\juyt{\eqalign{&\oint_{|z|=x^2}\frac{d z}{2\pi i z} \
E\big(-z^{2r-2}
x^{4rl+(4r-6)(r-1)}, x^{8(r-1)r}\big)
\ \frac{E(z x^{2r-3-2rl}, x^{2r})}{E(z^{-1} x^3,x^2)}=\cr
&\oint_{|z|=x^2}\frac{d z}{2\pi i z} \
z^{-2l}x^{6l}
E\big(-z^{2r-2}
x^{-4rl+(4r-6)(r-1)},x^{8(r-1)r} \big)
\ \frac{E(z x^{2r-3-2rl}, x^{2r})}{E(z^{-1} x^3,x^2)}\ .}}
To prove the equality we should
change  the
variable\ $z\to z^{-1}$ \ and
then  deform the integration
contour
at the left hand side of \juyt .

\listrefs
\end